\documentclass[11pt, a4paper]{article}

\usepackage{authblk}
\usepackage[colorlinks, citecolor=blue, linkcolor=blue, backref=true]{hyperref}
\usepackage{amssymb}
\usepackage{amsmath}
\usepackage{color}
\usepackage{enumerate}
\usepackage{amsthm}
\theoremstyle{theorem}

\usepackage{graphics,graphicx}

\author[1]{M.J. Nieves}
\author[2]{B.L. Sharma}
\affil[1]{\small Keele University, School of Computing and Mathematics, Keele, ST5 5BG, UK}
\affil[2]{\small Department of Mechanical Engineering, Indian Institute of Technology Kanpur, Kanpur, UP, 208016, India}
\date{}

\begin{document}
\title{{Interaction of in-plane waves with a structured penetrable line defect in an elastic lattice}}
\maketitle

\begin{abstract}
We consider the scattering of in-plane waves that interact with an edge of a structured {penetrable (inertial)} line defect contained in a triangular lattice, composed of periodically placed masses interconnected by massless elastic rods. The steady state problem for time-harmonic excitation is converted into a vector Wiener-Hopf equation using Fourier transform. The matrix Wiener-Hopf kernel of this equation describes all dynamic phenomena engaged in the scattering process, which includes instances where localised interfacial waves can emerge along structured defect. This information is exploited to identify the dependency of the existence of these waves on the incident wave parameters and properties of the inertial defect. The symmetry in the structure of scattering medium allows us to convert the vectorial problem into a pair of scalar Wiener-Hopf equations posed along the lattice row containing the defect. The solution embodies the exact representation of scattered field, in terms of a contour integral in the complex plane, that includes the contributions of evanescent and propagating waves. The solution reveals that in the remote lattice, the reflected and transmitted components of incident field are {accompanied by dynamic modes from three distinct symmetry classes in addition to localised interfacial waves}. These classes correspond to tensile modes acting transverse to the defected lattice row, shear modes that act parallel to this row, and wave modes represented as a mixture of these two responses. Benchmark finite element calculations are provided to validate results against the obtained semi-analytical solution, which involve numerical computations of the contour integrals. Graphical illustrations demonstrate special dynamic responses encountered during the wave scattering process, including dynamic anisotropy, negative reflection and negative refraction.
\end{abstract}


\section*{Introduction}

Scattering theory is the key to understanding a range of physical phenomena that occur when waves propagating through a uniform medium interact with an {obstacle \cite{FelsenMarcuvitz, Achenbach,YangAchenbach}}. The application of wave scattering problems is wide within science, engineering and technology; for example, they have been exploited to enhance detection methods in the non-destructive evaluation of materials and engineering structures \cite{Blitz,Fergusonetal}, {understanding physics of layered media \cite{Garnierbook, Garnier1, Gibson2, Gibson3}}, controlling and minimising noise in the natural environment \cite{Crighton,Zhuetal}, food processing and safety monitoring \cite{McClements,Pinfield}, coastal engineering \cite{AshokManam, Smithetal,Wilksetal, FMontieletal1}, amongst much else \cite{Sheng,Martin1,Colton}.

In a majority of the aforementioned applications, the medium supporting the waves is treated as continuous and the physical field of interest is either scalar or reducible to this form through some appropriate decomposition. In connection with this, there is a rich history of canonical scattering problems that provide building blocks for understanding wave scattering in more complex applications. 
These canonical problems include, for instance, the scattering of waves by a single defect and external and internal boundaries \cite{Vinogradovaetal}. 
In fact, the set of ``canonical diffraction problems'' is quite large and includes scattering of waves on infinite domains, domains with boundaries having corners and cusps, and therefore involve various singularities in the scattered field. Arguably, in this list, the most famous canonical diffraction problem is the ``Sommerfeld half-plane problem'', the focus of Sommerfeld's groundbreaking work
in the late 19th century \cite{Sommerfeld}. 

Sommerfeld presented the first analytical solution of the Helmholtz equation in 2D for a scalar field that satisfies either the Dirichlet or Neumann condition on a half-line contained in an infinite plane. 
One physical analogy of this problem includes an acoustic medium \cite{Morse} in three dimensions containing either a semi-infinite sound hard or sound soft screen interacting with a time harmonic incident plane wave. 
Mathematically, the screen is represented by a one-dimensional semi-infinite line in a two-dimensional continuum. This construct naturally introduces a discontinuity in conditions posed along the line containing the screen. In such problems, the Wiener-Hopf technique \cite{Noble} has played a key role in the derivation of their solution, leading to fields 
which may have certain 
singularities in the vicinity of the screen edge. The diffraction of plane elastic and electromagnetic waves by screens {and other obstacles} has been revise{studied in several classical works and continues to be an active area of research \cite{KuoYu}}. 
The analysis of the associated fields engage vectorial problems, where components of the field are coupled. However, as mentioned above, their solutions avail themselves to the analysis of scalar equations, through some appropriate decomposition involving scalar potentials. {Those problems often lead to the analysis of the scattering of a wider range of wave phenomena, e.g. in vector elasticity, where one can consider the scattering of surface or interfacial modes in addition to bulk waves. Surface modes are particularly important in applications such as seismic protection. The scattering of Rayleigh waves on an elastic half-space by near surface cracks has been tackled through the superposition principle and model problems involving wave radiation due to appropriately chosen forces in \cite{Wangetal}. An asymptotic analysis of transient wave scattering processes for Rayleigh waves produced by collections of resonators on the surface of an elastic half-space has been considered in \cite{ArgatovSabina}. For the scattering of elastic waves by inetrfaces between conventional and complex materials, see also \cite{Mokhtarietal}.

In the context of vectorial wave scattering problems, one can also consider the interplay of various physical fields. The coupling of sound and electric fields in wave scattering processes has been investigated in \cite{Markovetal} for acoustic waves interacting with with inclusions in porous media with electrolyte filled pores, analogous to scenarios found in petrophysics, and for poroelastic inclusions suspended in fluids, as commonly occurring in problems physical chemistry \cite{Markovetal2}.}

Having an immediate relevance to other problems encountered in optics and electromagnetism, and underlying links with different scattering problems considered there, the Sommerfeld half-plane problem has also prompted alternative mathematical derivations and extensions \cite{Achenbach, Kanaun} 
and, recently, generalizations to {discrete media in \cite{Bls1,Bls4a,Bls16}; see references therein for several classical works on Sommerfeld half-plane problem}. 

On the subject of wave propagation in discrete media, the set of ``canonical diffraction problems'' at par with those in continuum is also quite large. A detailed analysis of a discrete analogue of the Sommerfeld ‘soft’ screen and ‘hard’ screen problems, called `discrete Sommerfeld problems', has been presented in \cite{Bls1,Bls5}.
In contrast to the analysis of the scattering of waves in a continuum, the lattice brings new features including dispersion and dynamic anisotropy. Additionally, the elastic field in this problem is inherently regularized so it does not possess the classical singularity at the defect edge \cite{Bls4a, Bls4}, as expected in the analogous continuum problem \cite{HeinsSilver}. 
The mathematical approach has been extended to analyse discrete diffraction problems with different geometries \cite{Bls3,Bls2,Bls7,BlsM1,BlsM2}, waves interacting with rigid defects \cite{Bls5, Bls4,Bls3,Bls2}, finite length obstacles \cite{Bls4,Bls4a, Bls6}, and defects with damage zones \cite{Bls8}.

The paradigm of discrete periodic structures dates back, at least, to last couple of centuries \cite{Brillouin} {and lately finds interest in models based in machine learning \cite{Mortazavi2020, Ladygin2020,Garland2021, Chatterjee2021} and bio-membranes \cite{Perotti1}}.
{Lattice models are well established in the understanding the phenomena associated with dislocations \cite{Celli,Kresse,Bls016,Sharma2021}, phase boundaries and bi-stable structures \cite{Suzuki1981,Pouget1992,Abeyaratne,Puglisi,Truskinovsky03, sharma07,Truskinovsky2005,Cohen2014,Deng2020,Nitecki,Hwang,Horn,Chindam}.}
Wave propagation in discrete media has gained much attention in the last 20 years within physics, engineering and mathematics. This is partly due to the emergence of metamaterials that support counter-intuitive static and dynamic phenomena with extremely useful applications \cite{Zheludev,Kadic,Khuong2004,Nash2015,Mitchell2018,Wang2015,Zhang2020,Banerjee2019,Zhu2019,Nassar2020,Kim2023,Pajunen2021,Wang2019,Mueller2019,Palermo2022}. 
The theory of discrete media has been explored in several examples \cite{ScarpettaTibullo,Bacigalupo2021} including media capable of supporting localisation and lensing \cite{Colquittetal1}, negative refraction \cite{Tallericoetal, lawrieetal}, cloaking for bulk and surface waves \cite{Colquittetal2, Kadicetal, Coutantetal, Garauetal2019}, highly localised bulk waves \cite{CartaJonesetal}, uni-directional Stoneley and Rayleigh waves \cite{Garaueal, Nievesetal2020}, wave transmission in nano-ribbons and nano-tubes \cite{Bls6a,Bls6d} and surface wave scattering by interfaces \cite{Bls9,Blsv1,Blssurfinhomg}, effects of local resonators \cite{Huang2021, Huang2021a, Huang2023}, etc. Further, the exploitation of prestressed lattice structures and their effective behaviour has enabled the design of materials with tunable extremely localised elastic deformations \cite{Bordigaetal1}, elastic media with flutter instabilities \cite{Bordigaetal2} and bio-inspired systems \cite{BolshakRyvkin} that soften and absorb energy to remain intact when damaged. Lattice models can also be used as efficient tools to describe wave motion in materials at higher frequencies, capturing microstructural effects that their continuum counterparts cannot \cite{slepyansbook,ScarpettaTibullo}. Within the classical low-frequency regime they easily lead to effective partial differential equations accurately describing their behaviour \cite{Bls17}. Discrete models have been shown to produce effective partial differential equations in high frequency regimes \cite{Crasteretal2010} that enable one to predict the envelope function for highly anisotropic dynamic lattice responses.

In the aforementioned references, the wavefields are derived without knowledge of solid or structure's ability to sustain the considered waveforms. Hence, understanding how these special waveforms influence the material integrity is paramount for their application in the real-world \cite{Marder2010}. 
In this regard, we also mention the work of Slepyan in \cite{slepyansbook}, who analysed dynamic failure propagation in elastic discrete structures in order to explain the role of multi-scale processes in material fracture \cite{MarderGross, Bls016}.
The approach in \cite{slepyansbook}, developed for Mode III fracture of an square-cell elastic discrete structures, has also been extended to handle various dynamic failure phenomena in other {discrete structures \cite{Marder2015,Zhao2020,Chen2011}, including dissimilar systems \cite{Mishurisetal,BerinskiiSlepyan, GorbushinMishuris, Piccolroazetal}, lattices with structured interfaces \cite{Mishurisetal2009} and discrete chains with non-local interactions \cite{GorbushinMishuris2017}. It has also been extended to treat failure provoked by different loads, e.g. shearing \cite{Mishurisetal2012} and moving sources \cite{GorbushinMishuris2018}, occuring under different fracture criteria \cite{Gorbushinetal2017} and through different failure processes \cite{Mishurisetal2008}.} 
Further, models of flexural discrete structures have been analysed in \cite{Brunetal2013, Brunetal2014, Movchanetal2015, Nievesetal2016,Nievesetal2017, KamotskiSmyshlyaev, CherkaevRyvkin1, CherkaevRyvkin2}, which have substantial importance in understanding the vibration and collapse of engineering structures such as buildings and bridges. 

Besides its application to lattice vibrations, the mathematical theory of discrete scattering problems \cite{Shaban} also appear in the context of the discrete Schr\"{o}dinger equation {and analogous problems in graphs and networks} \cite{Delourme2023,Fefferman2022,Fefferman2022a,Zheng2019,Berkolaiko,Kuchment}. The discrete framework has been recently used to obtain exact solution of certain problems of electronic transport in ribbons and carbon nano-tubes \cite{Bls6b,Bls6c} across special junctions \cite{Kosynkin,Bachtold}.
More recently, analytical approaches for studying wave scattering in discrete media have appeared for structured media with more complex boundaries and techiques \cite{ShaninKorolkov,Shanin1} and random media \cite{Garnier}.
There are also several developments in the inverse problems for discrete media; see, for example, \cite{Isozaki}. A recent result on inverse scattering on lattices is given in \cite{Bls23,Bls23n2}; see references therein for inverse scattering in the discrete framework. 
From the viewpoint of multiple scattering, due to a large, possibly random, set of finite length defects of varying sizes, an alternative treatment is required \cite{Martin1,Garnier, Maurel,Abrahams,Garnierbook}.

In a majority of the lattice problems mentioned above, the fields of interest are solutions to scalar problems. Much less has been carried out in the literature in the case of vectorial problems, if one compares to a large volume of results in the continuum theory of elasticity. 
Mode I and II fracture of a triangular {lattice} {was considered in \cite{Kulakhmetova,Murat,Monette1994,MarderGross,Slepyan2001triangular,Pechenik2002,Nievesetal2013}.}
{Recently the lattice models have become relevant for several dynamical problems including edge mode propagation, see, for example, \cite{Lee2018,Chen2019}}.
Rayleigh wave propagation along the surfaces of gyro-elastic media was treated in \cite{Nievesetal2020, Nievesetal2021}. Lamb wave propagation in elastic strips has been analysed in \cite{Nievesetal2023}, with the theory recently being generalised to design gyro-elastic network capable of one-way wave propagation along pre-determined wave transport paths \cite{Cartaetal2023}. Further, the dynamic motion of origami-inspired structures has been considered in \cite{BerinskiiEremeyev}.
The current study is a part of researches \cite{Blstgnew} that build on this work by analysing, for the first time, discrete in-plane wave scattering by a semi-infinite {penetrable} (inertial) defect in an elastic structure.

\begin{figure}[!ht]
\center{\includegraphics[width=1\textwidth]{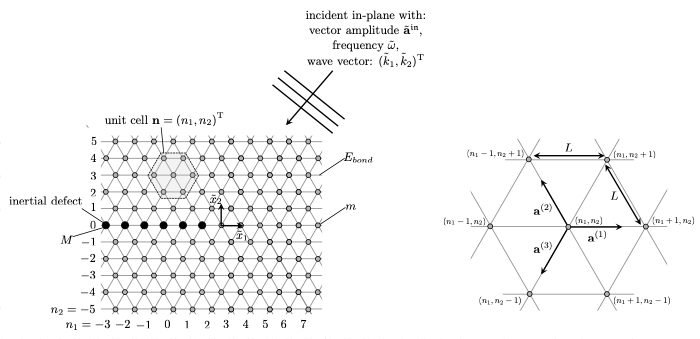}}

~~~~~(a)~~~~~~~~~~~~~~~~~~~~~~~~~~~~~~~~~~~~~~~~~~~(b)
\caption{(a) An elastic lattice containing an inertial linear defect interacting with a in-plane incident wave. (b) The interaction-cell corresponding to ${\bf n}=(n_1,n_2)^{\rm T}$ that is indicated by the shaded hexagon in (a). Note, depending on the choice of ${\bf n}$, this cell may contain one or at most three nodes belonging to the inertial defect.}
\label{pic1}
\end{figure}

We focus on a new paradigm involving the scattering of in-plane elastic waves by a semi-infinite defect. 
The theory of wave scattering in elastic discrete periodic structures containing linear defects was initiated in \cite{Bls1,Bls4a},where scattering of waves in an infinite square-cell lattice by a crack was investigated. 
In our work, the scattering problem is posed on a similar discrete medium but allowing for in-plane motion, which leads to a vectorial problem, and, therefore, brings out a distinct flavor and challenges. The assumed discrete medium is a triangular lattice formed from periodically placed masses, interconnected by massless elastic rods, and an embedded inertial defect; see Fig. \ref{pic1}(a). In the absence of a suitable decomposition of the solution into scalar potentials typically used in solving the continuum problem, we develop a method that handles the full vectorial nature of the problem. We demonstrate that the lattice problem is reducible to a scalar Wiener-Hopf equation due to certain symmetry in the assumed structure of triangular lattice with inertial defect.
This is used to reveal dynamic effects produced by the defect's presence, including anisotropy in its scattering response, support of interfacial waves enhancing wave motion near the defect and negative reflection and refraction. 
Note, with the examples mentioned in \cite{Bls8, Blsmatrix1,Blsmatrix2,Sharma2019wh}, even in case of scattering of scalar waves in discrete medium, it may happen that the formulation of \cite{Bls1} leads to a matrix Wiener-Hopf problems whose general solution remains an open question \cite{DAetal}. This makes the construction of scattering solutions more difficult in such situations while involving some technical difficulties. An extension of our problem does lead to a similar situation and it is currently under investigation with a matrix Wiener-Hopf kernel.

As in \cite{Bls1}, the solution to the discrete scattering problem is derived using the `discrete' Fourier transform (also known as $Z$-transform) to reduce the governing equations to a Wiener-Hopf equation posed along the line defect. In the application of the discrete Wiener–Hopf method (equivalent to Toeplitz operator based analysis \cite{Bls4a,Bls4,Bls6}), the equation includes the so-called kernel function that embeds all information about possible waves in lattice produced by the defect interacting with incoming waves. In our problem, after exploiting a symmetry, the Wiener-Hopf equation is reduced to a scalar form, similar to most cases mentioned above, whose general solution is well-known and readily obtained with methods from complex analysis. The subsequent solution is used to reproduce the dynamic behaviour of the system through the inverse Fourier transform. 
The scattering response of the defect begins to assume non-trivial wave forms in high-frequency regimes. 

The details are provided in the paper where we demonstrate the difference in response for defects having smaller inertia (Figure \ref{pic_beta_m0p5}(a)) vis-a-vis greater inertia (Figure \ref{pic_beta_m0p5}(b)) than the ambient lattice. 
We note that these features discussed are consistent with the slowness contours for the bulk medium in Figure \ref{pic_beta_m0p5}(c), which in this particular case are similar to those encountered in structured materials that exhibit strong dynamic anisotropy and localised waves.
Overall the discreteness of the medium in this vectorial framework brings out features which are also consistent with other problems in scalar case and continuum models. 

The structure of the article is as follows. Section \ref{sec2} describes the problem of in-plane wave scattering in an elastic lattice with an inertial defect. Section \ref{sec3} contains the governing equations for the lattice system, and their solution via the Fourier transform which leads to a Wiener-Hopf equation posed along the line containing the defect.
Section \ref{sec4} is devoted to the analysis of waves propagating in the lattice during the scattering process. This section also contains the analysis of the dispersion relations of considered system and the description of the corresponding lattice waves.
In particular, we discuss regimes where the lattice defect can support localized waves.
In Section \ref{sec5}, we provide the benchmark comparison of the solution to the problem via the Wiener-Hopf technique with computations performed using finite elements. We also describe the dynamic behaviour of the system for various inertial defects and frequencies, with the aid of numerical computations based on the analytical solution. Finally, in Section \ref{sec6} we give some conclusions and discussion.

\section{Formulation of scattering due to an inertial defect in two dimensional lattice}\label{sec2}

{{\bf \emph{The triangular lattice.}}}
We use the notation ${\bf n}=(n_1, n_2)^{\rm T}\in{{{\mathbb{Z}}}^2}$, to describe the positions of nodes in the medium schematically shown in Figure \ref{pic1}(a), where ${{\mathbb{Z}}}$ denotes the set of integers, ${{{{\mathbb{Z}}}^2}}$ denotes ${{\mathbb{Z}}}\times{{\mathbb{Z}}}$. The associated infinite triangular lattice is composed of periodically placed point masses, interconnected by massless elastic rods of stiffness ${E}$ and undeformed length $L$.
In its undeformed configuration, the physical structure belongs to the plane $\mathbb{R}^2$ and the positions of its nodes coincide with elements of the set
\[
\mathcal{L}:=\bigg\{\bigg((n_1+n_2/2)L, \sqrt{3}n_2 L/2\bigg)^T: n_1, n_2\in\mathbb{Z}\bigg\}\subset\mathbb{R}^2,
\]
{with $\mathbb{R}$ denoting the real numbers.}

During in-plane motion of the structure, the subsequent configurations belong to the same $\mathbb{R}^2$.
With each node {\bf n} in the triangular lattice, we associate an interaction-cell shown in Figure \ref{pic1}(b). We consider the interactions {between} the node ${\bf n}$ {and} its nearest neighbors, enumerated in accordance with the notation depicted in Figure \ref{pic1}(b). 

{{\bf \emph{The inertial line defect.}}} 
Let ${{\mathbb{Z}}_+}$ denote the set of non-negative integers
and ${{\mathbb{Z}}_-}$ denote the set of negative integers.
The lattice nodes corresponding to $n_1\in{{\mathbb{Z}}_-}, n_2=0$ have mass $M=m(1+\beta)$, whereas as all other nodes have mass $m$, with $\beta\ge -1$.
According to this, if $\beta \ne 0$ the nodes with mass $M$ form a linear inertial defect in the lattice and we define several classes of inertial defects:

\begin{enumerate}[$\bullet$]
\item $-1<\beta<0$ describes a ``light" inertial defect, i.e. whose inertia is smaller than that of the ambient medium.
\item $\beta>0$ indicates the presence of a ``heavy" inertial defect.
\item $\beta=-1$ is a special case and corresponds to a non-inertial defect, where there are no point masses located at the junctions described by the indices $n_1\in{{\mathbb{Z}}_-}, n_2=0$.
\end{enumerate}
{Note if $\beta=0$, the lattice is uniform and therefore there is no question of wave scattering. In fact, the corresponding problem can be interpreted as the one relevant for the description of wave incident on the inertial defect (see Section \ref{inc_w_sec}). }

\vspace{0.1in}This paper is concerned with the influence of the inertial defect on the lattice waves when interacting with an incident wave.

{{\bf \emph{The total
field, scattered field, and dimensionless variables.}}} 
The physical lattice displacement is represented by the vector field ${\tilde{\bf u}}: \mathcal{L}\times\mathbb{R}\to{\mathbb{R}}^2$; thus, it is dependent on the physical location of nodes and time.
The dimensionless total displacement ${{\bf u}}^{\mathfrak{t}}$ {of the point mass at node ${\bf n}$} is related to ${\tilde{\bf u}}$ according to 
\begin{equation}
{{{{\tilde{\bf u}}}}(\tilde{x}_1, \tilde{x}_2, \tilde{t})}={{\bf u}}^{\mathfrak{t}}_{{\tt x}, {\tt y}}(t){L},
\label{un1n2}
\end{equation}
where the physical space coordinates of the nodes $\tilde{x}_j$,$j=1,2$, and time $\tilde{t}$ are related to their dimensionless variables $n_1,n_2$ and $t$, respectively, by
\begin{equation}
\tilde{x}_1=(n_1+n_2/2)L, \quad \tilde{x}_2=\sqrt{3}n_2 L/2, \quad {\tilde{t}}= t\sqrt{m/{E}}.
\label{x1x2n1n2}
\end{equation}
Recall that $L$ is the undeformed lattice spacing (see Figure \ref{pic1}).
Following \cite{Bls1}, in place of $\tilde{x}_1, \tilde{x}_2$ or $n_1, n_2$, it is more convenient to use the dimensionless lattice coordinates $({\tt x}, {\tt y})$ whose relationship with $n_1, n_2$ is defined by
\begin{equation}
{\tt x}:=2n_1+n_2\;, {\tt y}:=n_2\;,\qquad n_1, n_2\in \mathbb{Z}.
\label{xyn1n2}
\end{equation}
As a result of \eqref{x1x2n1n2}, $\tilde{x}_1={\tt x}L/2, \tilde{x}_2=\sqrt{3}{\tt y} L/2$. Later we also use $x_j=\tilde{x}_j/L$, $j=1,2$, to denote the dimensionless principal coordinates of the system (see Section \ref{sec_disp_bulk} onwards). In the following, we often write $\sum_{n_1\in\mathbb{Z}}$ at a given $n_2$ where the summand is a function of ${\tt x}, {\tt y}$ according to above relationship \eqref{xyn1n2}; in particular, ${\tt x}$ takes even (resp. odd) integer values for even (resp. odd) values of ${\tt y}$ for nodes in the physical triangular lattice.

It is emphasized that the dimensionless vector valued function ${{\bf u}}^{\mathfrak{t}}_{{\tt x}, {\tt y}}(t)$
in \eqref{un1n2}, in combination with \eqref{xyn1n2}, is used to represent the in-plane {\em total displacement {field}} 
for the structured medium. It is a tacit assumption that there is no out-of-plane displacement of the nodes in the lattice as part of three dimensional ambient space.
Indeed, the standard {additive composition} of ${{\bf u}}^{\mathfrak{t}}_{{\tt x}, {\tt y}}(t)$ is in terms of the incident field ${{\bf u}}^{\mathfrak{i}}_{{\tt x}, {\tt y}}(t)$ and scattered field ${{\bf u}}^{\mathfrak{s}}_{{\tt x}, {\tt y}}(t)$ as follows:
\begin{equation}
{{\bf u}}^{\mathfrak{t}}_{{\tt x},{\tt y}}(t)={{\bf u}}^{\mathfrak{i}}_{{\tt x}, {\tt y}}(t)+{{\bf u}}^{\mathfrak{s}}_{{\tt x}, {\tt y}}(t).
\label{incscasum}
\end{equation}

The incident wave is known apriori and is assumed to be time harmonic. It is more convenient to consider the incident, scattered and total fields as vector valued displacements with complex entries {in $\mathbb{C}$}. 
It is henceforth assumed that
\begin{equation}\label{uinc}
{{\bf u}}^{\mathfrak{i}}_{{\tt x}, {\tt y}}(t)={\bf a\rm}^{\mathfrak{i}} e^{{\rm i}(k_{\tt x} {\tt x}+k_{\tt y}{\tt y}-\omega t)}\;,
\end{equation}
where ${\bf a\rm}^{\mathfrak{i}}\in{\mathbb{C}}^2$ is the incident wave {polarization and} amplitude, 
$\omega\in\mathbb{R}$ is the dimensionless radian frequency and $k_{\tt x}\in\mathbb{R}$ and $k_{\tt y}\in\mathbb{R}$ are the incident wave's dimensionless wavenumbers {along the lateral and vertical lattice directions (see also \eqref{x1x2n1n2} for physical coordinates)}, respectively. 

The counterparts of the parameters in \eqref{uinc} defining the incident wave's temporal and spatial behaviour in the physical space are then 
\begin{equation}\label{dim5}
\tilde{k}_1=2k_{\tt x}/L\;, \quad \tilde{k}_2=2k_{\tt y}/(\sqrt{3}L), \text{ and } \tilde{\omega}={\omega} \sqrt{{E}/m}.
\end{equation}
Note that the physical frequency $\tilde{\omega}$ and wavenumbers $\tilde{k}_1$ and $\tilde{k}_2$ are interconnected and their relationship is found through the analysis of the homogeneous dynamic equations in the bulk of the medium {(i.e., away from $n_2=0$)}. We explore this connection later for their dimensionless equivalents. 
Knowing the relationship between these parameters then enables one to obtain the eigenvector ${\bf a\rm}^{\mathfrak{i}}$ and fully define the incident wave. 

In the considered elastic lattice problem the incident field \eqref{uinc} can be represented either by a shear wave, a pressure wave or a combination of these waves. However, despite an intrinsic anisotropy it is possible in the assumed lattice model to restrict our attention to the case where the incident field is one of these two types of waveforms.

In the context of \eqref{incscasum}, in what follows our goal is to determine the scattered field ${{\mathbf{u}}}_{{{\tt x}, {\tt y}}}$ produced by the incident field \eqref{uinc} in {steady state regime}. Thus it is natural to seek the time dependent scattered field ${{\bf u}}_{{\tt x}, {\tt y}}^{\mathfrak{s}}$ of the form 
\begin{equation}\label{Us}
{{\bf u}}_{{\tt x}, {\tt y}}^{\mathfrak{s}}(t)={{\mathbf{u}}}_{{{\tt x}, {\tt y}}}e^{-{\rm i}
\omega t}.
\end{equation}
where ${{\mathbf{u}}}_{{{\tt x}, {\tt y}}}\in{\mathbb{C}}^2$ is time-independent. 

{{\bf \emph{The limiting absorption principle and complex wave parameters}}
The rigorous argument justifying the steady state is technically referred as the limiting absorption principle, equivalent to hypothesis of radiation conditions $\cite{Shaban}$, and this involves a limit of the scattered field obtained by assuming a complex value of $\omega$ such that $\text{{\rm Im}}(\omega)\to0+$. Naturally, as a consequence of this, the analysis involves complex values of $k_{\tt x}$ and $k_{\tt y}$ as well. Thus, the incident wave is described by parameters $\omega, k_{\tt x}, k_{\tt y}\in\mathbb{C}$ with vanishing imaginary parts. We abuse the notation, for convenience, and ignore the imaginary parts in most of the discussion.}

\section{{Equations of the lattice dynamics}}\label{sec3}
For $n_2\in\mathbb{Z}\setminus\{0\}$, the dimensionless nodal displacements \eqref{un1n2} satisfy the linear momentum balance law
\begin{eqnarray}
 \ddot{{\bf u}}^{\mathfrak{t}}_{{\tt x}, {\tt y}}&=& {\bf a}^{(1)}\cdot \left[{\bf u}^{\mathfrak{t}}_{{\tt x}+2, {\tt y}}+{\bf u}^{\mathfrak{t}}_{{\tt x}-2, {\tt y}}-2{\bf u}^{\mathfrak{t}}_{{\tt x}, {\tt y}}\right]{\bf a}^{(1)}\nonumber \\
&&+{\bf a}^{(2)}\cdot \left[{\bf u}^{\mathfrak{t}}_{{\tt x}-1, {\tt y}+1}+{\bf u}^{\mathfrak{t}}_{{\tt x}+1, {\tt y}-1}-2{\bf u}^{\mathfrak{t}}_{{\tt x}, {\tt y}}\right]{\bf a}^{(2)}\nonumber\\
&&+{\bf a}^{(3)}\cdot \left[{\bf u}^{\mathfrak{t}}_{{\tt x}+1, {\tt y}+1}+{\bf u}^{\mathfrak{t}}_{{\tt x}-1, {\tt y}-1}-2{\bf u}^{\mathfrak{t}}_{{\tt x}, {\tt y}}\right]{\bf a}^{(3)}\;,\label{eq2a}
\end{eqnarray}
where the vectors ${\bf a}^{(j)}$, $1\le j \le 3$, used above are
\begin{equation}
{\bf a}^{(1)}=(1, 0)^{\rm T}\;, \quad {\bf a}^{(2)}=(-1/2, \sqrt{3}/2)^{\rm T}\;, \quad {\bf a}^{(3)}=(-1/2, -\sqrt{3}/2)^{\rm T},
\end{equation}
directed as in Figure \ref{pic1}(b). {We define ${\bf a}\cdot{\bf b}=a_1a_2+b_1b_2$ for ${\bf a}=[a_j]_{j=1}^2, {\bf b}=[b_j]_{j=1}^2\in {\mathbb{R}}^2$; for example, (\ref{eq2a}) utilises this inner-product. We also use this notation later for the dot product of vectors with complex entries.} Additionally, the dot over the displacement in (\ref{eq2a}) represents the derivative with respect to time.

As mentioned above, the lattice row at $n_2=0$ is partially occupied by an inertial defect, located at $n_1\in{{\mathbb{Z}}_-}$. The equations describing the motion of the nodes along the row $n_2=0$ are:
\begin{eqnarray}
\ddot{{\bf u}}^{\mathfrak{t}}_{{\tt x}, 0}H({\tt x})
+(1+\beta) \ddot{{\bf u}}^{\mathfrak{t}}_{{\tt x}, 0} 
H(-1-{\tt x})&=& {\bf a}^{(1)}\cdot \left[{\bf u}^{\mathfrak{t}}_{{\tt x}+2, 0}+{\bf u}^{\mathfrak{t}}_{{\tt x}-2, 0}-2{\bf u}^{\mathfrak{t}}_{{\tt x}, 0}\right]{\bf a}^{(1)}\nonumber \\
&&+{\bf a}^{(2)}\cdot \left[{\bf u}^{\mathfrak{t}}_{{\tt x}-1,1}+{\bf u}^{\mathfrak{t}}_{{\tt x}+1, -1}-2{\bf u}^{\mathfrak{t}}_{{\tt x}, 0}\right]{\bf a}^{(2)}\nonumber\\
&&+{\bf a}^{(3)}\cdot \left[{\bf u}^{\mathfrak{t}}_{{\tt x}+1, 1}+{\bf u}^{\mathfrak{t}}_{{\tt x}-1, -1}-2{\bf u}^{\mathfrak{t}}_{{\tt x},0}\right]{\bf a}^{(3)}\;,\label{eq2b}
\end{eqnarray}
where the factor $1+\beta$ is the distributed mass along the inertial defect (see Section \ref{sec2}) and $H(x)$ is the discrete Heaviside function:
\begin{equation}
H(x):=\left\{\begin{array}{ll}
1\;, & \quad \text{ if } x \in{{\mathbb{Z}}_+}\\
0\;, &\quad \text{ if } x \in{{\mathbb{Z}}_-}.
\end{array}\right.
\label{Heavi}
\end{equation}
Note that the equations corresponding to the motion of the lattice in the physical space can be derived from (\ref{eq2a}) and (\ref{eq2b}) in terms of quantities involving dimensions through the definitions and normalisations in (\ref{un1n2})--(\ref{dim5}).

\subsection{Propagation of the incident wave within the lattice}\label{inc_w_sec}
There is a certain admissible range of parameters in \eqref{uinc} such that the lattice microstructure exterior to the inertial defect can support the incident wave. Defining the incident and using the equations of the previous section helps to determine the scattered field in the steady state regime.
 
\vspace{0.1in}{\bf \emph{Admissible incident waves.}} We require that the incident field is a solution of homogeneous steady state analogue of (\ref{eq2a}). Thus taking the vector function \eqref{uinc} representing the incident wave 
(see (\ref{uinc})) and inserting this into (\ref{eq2a}) gives the homogeneous algebraic system:
\begin{equation}\label{ID}
(\omega^2 {\bf I}-{\bf A}({k}_{\tt x},{k}_{\tt y})) {{\bf a\rm}}^{\mathfrak{i}}={\bf 0}\;,
\end{equation}
where {${\bf I}$ is the $2\times 2$ identity matrix and }
\[{\bf A}({k}_{\tt x},{k}_{\tt y})=\left(\begin{array}{cc}
3-2\cos(2k_{\tt x})-\cos(k_{\tt x})\cos(k_{\tt y}) & {\sqrt{3}}\sin(k_{\tt x})\sin(k_{\tt y}) \\ \\
 {\sqrt{3}}\sin(k_{\tt x})\sin(k_{\tt y}) & {3}(1-\cos(k_{\tt x})\cos(k_{\tt y}) )
\end{array}\right)\;.\]
It follows that a non-trivial incident wave can propagate in the medium if the coefficient matrix in (\ref{ID}) is degenerate. This occurs provided the wavenumbers ${k}_{\tt x}$ and ${k}_{\tt y}$ are related to the incident wave radian frequency $\omega$ through one of the following expressions:
\begin{equation}\label{incdisp}
\omega_\pm({k}_{\tt x},{k}_{\tt y}):=\sqrt{\frac{(\text{tr}{\bf A})\pm\sqrt{{(\text{tr}{\bf A})^2-4\text{det}({\bf A})}}}{2}},
\end{equation}
where ${\bf A}\equiv {\bf A}({k}_{\tt x},{k}_{\tt y}).$ The functions $\omega_\pm({k}_{\tt x},{k}_{\tt y})$ provide the dispersion surfaces for the uniform lattice structure (without defect i.e. $\beta=0$). They can be further classified into acoustic and optical bands as presented later in the paper.

\begin{figure}[!ht]
\center{\includegraphics[width=0.5\textwidth]{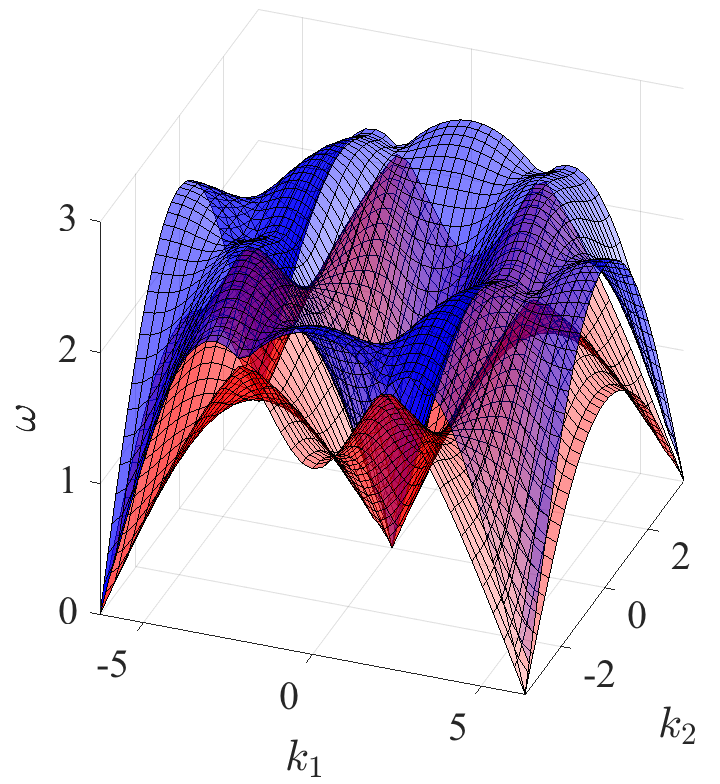}}
\caption{Dispersion surfaces $\omega_-$ (red) and $\omega_+$ (blue) as a function of $k_1, k_2$ for the infinite uniform triangular lattice based on (\ref{incdisp}) and using \eqref{k1k2}.}
\label{pic_disp_surf}
\end{figure}

\vspace{0.1in}{{\bf \emph{Long wavelength limit of the lattice and associated dispersion relations:}}
The surfaces corresponding to $\omega_{\pm}$ as functions of the dimensionless wavenumbers 
\begin{equation}\label{k1k2}
k_1:= 2{k_{\tt x}}\text{ and }k_2:=2{k_{\tt y}}/\sqrt{3}
\end{equation}
(see also $(\ref{dim5})$) are shown in Figure \ref{pic_disp_surf}. 
The corresponding wave vector is also defined by
\begin{equation}\label{k1k2vec}
{\bf k}:=(k_1, k_2)^T\in\mathbb{R}^2.
\end{equation}
The expressions $\omega_-$ and $\omega_+$ are connected with shear and pressure waves, respectively, propagating in the corresponding homogenized model for an elastic continuum when $|{\bf k}|\to 0$. Indeed, in the long wavelength limit the behaviour of bulk lattice corresponds to the dynamic plane strain deformation of an isotropic, homogeneous, linear elastic material $\cite{slepyansbook}$ with the effective Lam\'e parameters $\lambda_{\rm e}=\mu_{\rm e}=\sqrt{3}{E}/4$ and effective mass density $\rho_{\rm e}=2m/(\sqrt{3}L^2)$. Further, the pressure and shear wave speeds for this homogenised medium are $v_p=\sqrt{9/8}{E} L^2/m$ and $v_s=\sqrt{3/8}{E} L^2/m$, respectively. In addition, as $|{\bf k}| \to 0$ one can show that the two branches in \eqref{incdisp} behave
 in the following manner:
\begin{equation}
\omega_-= v_s m/({E} L^2)|{\bf k}|+o({\bf k}) \quad \text{ and } \quad\omega_+=v_pm/({E} L^2) |{\bf k}|+o({\bf k}),
\end{equation}
where the leading order terms in the expansion describe the dispersion relations obtained for the effective elastic continuum.
Such a continuum limit of the discrete scattering problem of this article can be carried in a manner similar to the case of discrete Sommerfeld problems \cite{Bls17,Bls4a}.}

\vspace{0.1in} {\bf \emph{Determining the incident wave.}} It can be determined from (\ref{pic_disp_surf}) that the admissible frequency range for the incident wave is $(0, \sqrt{6})$. 
An incident wave is defined by selecting a frequency $\omega$ that defines a horizontal plane intersecting these surfaces, and consequently slowness contours corresponding to where (\ref{incdisp}) are equal to the frequency. The outward facing normal vector associated with $\nabla_{\bf k} \omega_{\pm}$, then defines the group velocity of the incident wave. Here, where ${\bf k}$ (given in \eqref{k1k2}) and \eqref{k1k2vec}) describes its direction of propagation with respect to the principal coordinates of the lattice.
The {polarization} and amplitude of the incident wave is found through (\ref{ID}) as 
\begin{equation}\label{ainc}{{\bf a\rm}}^{\mathfrak{i}}={\rm a} \Big(1,
\dfrac{\sqrt{3}\sin(k_{\tt x})\sin(k_{\tt y})}{\omega^2-3(1-\cos(k_{\tt x})\cos(k_{\tt y}))}\Big)^T.\end{equation}
The constant ${\rm a}$ can be specified to scale the amplitude of the incident wave.

\subsection{Solution for the scattered field in the bulk lattice}\label{sec2p2}
We rewrite (\ref{eq2a}) and (\ref{eq2b}) in terms of the scattered field ${\mathbf{u}}_{{\tt x}, {\tt y}}$ using (\ref{Us}) and (\ref{incscasum}) and 
we apply the transformation ${\mathbf{u}}_{{\tt x}, {\tt y}}\to {\mathbf{u}}^{{\rm F}}_{{\tt y}}(z)$, at a fixed value of ${\tt y}$, according to the formula
\begin{equation}\label{ZT}
{\mathbf{u}}^{{\rm F}}_{{\tt y}}(z)=\sum_{{n_1}\in \mathbb{Z}}{\mathbf{u}}_{{\tt x}, {\tt y}} z^{-{\tt x}},
\end{equation}
where $z$ belongs to an appropriate annulus ${\mathcal{A}}$ in the complex plane so that the preceding infinite sum converges and in which the known functions contained in the Wiener-Hopf equation obtained later are analytic \cite{Bls1}.
In doing so, for the bulk lattice, corresponding to ${\tt y}\in\mathbb{Z}\setminus\{0\}$, the equations of motion (\ref{eq2a}) are transformed to 
\begin{eqnarray}
-\omega^2 {{\mathbf{u}}^{{\rm F}}_{{\tt y}}}&=& {\bf a}^{(1)}\cdot {\mathbf{u}}^{{\rm F}}_{{\tt y}}\left[z^2+z^{-2}-2\right]{\bf a}^{(1)}\nonumber \\
&&+{\bf a}^{(2)}\cdot \left[{\mathbf{u}}^{{\rm F}}_{{\tt y}+1}z^{-1}+{\mathbf{u}}^{{\rm F}}_{{\tt y}-1}z-2{\mathbf{u}}^{{\rm F}}_{{\tt y}}\right]{\bf a}^{(2)}\nonumber\\
&&+{\bf a}^{(3)}\cdot \left[{\mathbf{u}}^{{\rm F}}_{{\tt y}+1}z+{\mathbf{u}}^{{\rm F}}_{{\tt y}-1} z^{-1}-2{\mathbf{u}}^{{\rm F}}_{{\tt y}}\right]{\bf a}^{(3)}\;.\label{bulk_eq_tran}
\end{eqnarray}
It is natural assumption that the scattered wave field should decay away from the edge of defect.
As demonstrated in the Appendix
\ref{Uupper} the scattered field in the bulk lattice for ${\tt y}\in\mathbb{Z}\setminus\{0\}$ can be written in terms of its value at ${\tt y}={0}$ as
\begin{equation}\label{u_upper}
{\mathbf{u}}^{\rm F}_{{\tt y}}=\left\{\begin{array}{ll}
 {\bf B \rm}\Lambda^{{\tt y}} {\bf B}^{-1} \mathbf{u}^{\rm F}_0\;, & \quad \text{ if } {\tt y} \in{{\mathbb{Z}}_+}
\\ \\
{{{\textbf J}}}{\bf B \rm}\Lambda^{-{\tt y}} {\bf B}^{-1}{{\textbf J}} \mathbf{u}^{\rm F}_0\;,& \quad \text{ if } {\tt y} \in{{\mathbb{Z}}_-}\cup\{0\},
\end{array}\right.,
\end{equation}
where
\begin{equation}
{{\textbf J}}=\left[\begin{array}{ll}
1 & \,\,\,\, 0\\
0&-1
\end{array}\right]\;,\qquad
{\textbf B}=[{\textbf c}(\lambda_1)\;\; {\textbf c}(\lambda_2) ],\qquad{\textbf c}(\lambda)=\left[ \begin{array}{c} 1 \\ \\ \displaystyle{\frac{-\frac{\sqrt{3}}{4} (\lambda-{\lambda}^{-1})(z-{z}^{-1}) }{ \frac{3}{4}\, \left( \lambda+{\lambda}^{-1} \right) \left( z+{z}^{-1}
 \right) -3+ \omega^2}}\end{array}
\right],
\label{BM}
\end{equation}
and
\begin{equation}\begin{split}
\label{lam_sol}
\Lambda=\text{diag}\{\lambda_1, \lambda_2\}\;,
\lambda_j\equiv \lambda_j(z)=\frac{\sqrt{m_j(z)+1}-\sqrt{m_j(z)-1}}{\sqrt{m_j(z)+1}+\sqrt{m_j(z)-1}}\;, \quad j=1,2\\
m_1(z)=b(z)+\sqrt{(b(z))^2-c(z)}\;, \quad m_2(z)=b(z)-\sqrt{(b(z))^2-c(z)}\;,\\
b(z)=-\frac{1}{4}\left(z+z^{-1}\right)\left(\left(z+z^{-1}\right)^2-6+\frac{4}{3}\omega^2\right)\;,\\
\text{ and }
c(z)=\left(\frac{1}{3}\omega^2-\frac{3}{4}\right)\left(z+z^{-1}\right)^2+\frac{1}{3}\left(\omega^2-2\right)\left(\omega^2-6\right)\;.
\end{split}\end{equation}
In above, $z$ belongs to the annulus ${\mathcal{A}}$. {The square root function, $\sqrt{\cdot}$, has the usual branch cut in the complex plane running from $-\infty$ to $0$.}
{Note that $|\lambda_{1,2}|< 1$ if $\text{\rm Im}(\omega)>0$ and the remaining pair of roots $\lambda_{1,2}^{-1}$ of $(\ref{lam_bi_quad})$ do not contribute in $\eqref{u_upper}$ as their modulus exceeds unity and the scattered wave field is required to decay at infinity, within the framework of the limiting absorption principle.}
Based on \eqref{u_upper}, it is clear that after $\mathbf{u}^{\rm F}_0$ is determined, we can find the bulk lattice response in the considered scattering problem.

\subsection{The Wiener-Hopf equation}\label{sec2p3}
\label{sec4}
We focus on (\ref{eq2b}) posed along the lattice row embedding the inertial defect. Using the definition \eqref{Heavi} of the discrete Heaviside function, (\ref{eq2b}) can be rewritten in terms of the scattered field by employing (\ref{incscasum}), \eqref{uinc}, and (\ref{Us}) to obtain an equation involving the scattered field given by:
\begin{eqnarray}
&&-\omega^2 {{\mathbf{u}}}_{{\tt x}, 0}
-\omega^2 \beta {{\mathbf{u}}}_{{\tt x}, 0}H(-1-{{\tt x}})-\omega^2 \beta {{\bf a}}^{\mathfrak{i}}e^{\text{i}k_{\tt x} {\tt x}}H(-1-{{\tt x}})\nonumber \\
&=& {\bf a}^{(1)}\cdot \left[{\mathbf{u}}_{{\tt x}+2, 0}+{\mathbf{u}}_{{\tt x}-2, 0}-2{\mathbf{u}}_{{\tt x}, 0}\right]{\bf a}^{(1)}\nonumber \\
&&+{\bf a}^{(2)}\cdot \left[{\mathbf{u}}_{{\tt x}-1,1}+{\mathbf{u}}_{{\tt x}+1, -1}-2{\mathbf{u}}_{{\tt x}, 0}\right]{\bf a}^{(2)}\nonumber\\
&&+{\bf a}^{(3)}\cdot \left[{\mathbf{u}}_{{\tt x}+1, 1}+{\mathbf{u}}_{{\tt x}-1, -1}-2{\mathbf{u}}_{{\tt x},0}\right]{\bf a}^{(3)}\;.\label{eq2bSC}
\end{eqnarray} 
In the Appendix \ref{appWH}, we show that following the application of (\ref{ZT}) and engaging the half-transforms ${\mathbf{u}}_{0\pm}$ defined as:
\begin{equation}\label{split}
{\mathbf{u}}_{0\pm}(z)=\sum_{n_1\in\mathbb{Z}_\pm}{\mathbf{u}}_{\tt x, 0}z^{-\tt x}
\end{equation}
and the additive split
\begin{equation}\label{splitpm}
{\mathbf{u}}^{\rm F}_{0}(z)={\mathbf{u}}_{0+}(z)+{\mathbf{u}}_{0-}(z),
\end{equation}
that (\ref{eq2bSC}) can be converted into the following Wiener-Hopf equation:
\begin{equation}\label{EqWH1}
{\mathbf{u}}_{0+}(z)+{{\bf L}(z)}{\mathbf{u}}_{0-}(z)=({\bf I}-{\bf L}(z)){{\bf a}}^{\mathfrak{i}} \mathrm{f}_-(z),
\end{equation}
where $z\in{\mathcal{A}}$, with ${\mathcal{A}}$ being the annulus where entries of ${\bf L}$ and the above right-hand side is analytic (see also Section \ref{sec2p2}).
In relation to the above equation, the terms supplied with a ``$-$" (``$+$") represent functions analytic in the union of the region in $\mathbb{C}$ inside (outside) of the annulus ${\mathcal{A}}$ and ${\mathcal{A}}$ itself. 
Further, the matrix kernel ${\bf L}(z)$ in (\ref{EqWH1}) is in fact diagonal, which can be verified by direct computation and using the representation of $\lambda_j$ and $n_j$, $j=1,2$ (see (\ref{lam_sol})). The matrix kernel ${\bf L}(z)$ takes the simplified form
\begin{equation}\label{ML}
 {\bf L}(z)=\left[\begin{array}{cc}
\mathfrak{L}_1(z) & 0\\
0 & \mathfrak{L}_2(z)
\end{array}\right]
\end{equation}
with
\begin{eqnarray}
\mathfrak{L}_1(z)&=&1-\frac{2\beta\omega^2\lambda_1\lambda_2[3(\lambda_1+\lambda_2)(z^2+1)+2z(\omega^2-3)(1+\lambda_1\lambda_2)]}{3z(\lambda_1^2-1)(\lambda^2_2-1) (\lambda_1\lambda_2-1)}\;, \nonumber \\
\mathfrak{L}_2(z)&=&1+\frac{2\beta \omega^2 z[3(\lambda_1+\lambda_2)(z^2+1)+2z(\omega^2-3)(1+\lambda_1\lambda_2)]}{[9(z^2+1)^2-4z^2(\omega^2-3)^2](\lambda_1\lambda_2-1)}\;,\label{L1L2}
\end{eqnarray}
where $\lambda_1\equiv \lambda_1(z)$, $\lambda_2\equiv \lambda_2(z)$,
and (following \eqref{split} and \eqref{xyn1n2})
\begin{equation}
\mathrm{f}_-(z):=
\sum_{n_1\in\mathbb{Z}_-} e^{{\rm i}k_{\tt x}{\tt x}}z^{-{\tt x}}.
\label{Fm}
\end{equation} 
Here, $\mathrm{f}_-$ is the transform \eqref{ZT} of the incident field supported on the defect (that is $n_1\in\mathbb{Z}_-$). 

{The function $\mathfrak{L}_1(z)$ $($$\mathfrak{L}_2(z)$$)$ plays a role in characterising the longitudinal (transverse motion) of the lattice row at $n_2=0$, as well as the behaviour of the remaining lattice. This is discussed in detail later in the paper.}
{In view of the reduction to scalar Wiener-Hopf problems, the case of scattering due to finite length inertial defect admits an analysis following the lines of \cite{Bls4a,Bls4,Bls6}.}

{In view of statements in Section \ref{sec2p2}, if ${\rm Im}(\omega)>0$, the wave number $k_{\tt x}$ in (\ref{Fm}) is no longer real and possesses a small imaginary part. This results in the exponential term in embedded in the power series $(\ref{Fm})$ being located in the vicinity of the unit disk in $\mathbb{C}$.}

\subsection{The exact solution for the scattered field}
\label{secexactsol}
The solution to the problem of finding the complex functions ${\mathbf{u}}_{0\pm}$ in \eqref{EqWH1} can be obtained using the standard arguments from the Wiener-Hopf technique (see Appendix \ref{appWHsol} for the details).
Indeed, by using \eqref{splitpm}, the solution of (\ref{EqWH1}) leads to
\begin{equation}\label{SOL}
{\mathbf{u}}^{\rm F}_{0}(z)=[{\bf L}_-(z)]^{-1}(\mathbf{I}-\mathbf{L}(z))\sum_{j=1}^2 [{\bf L}_+(w_j)]^{-1}{\bf a}^{\mathfrak{i}}\mathrm{f}^{(j)}_-(z),
\end{equation}
where 
\begin{equation}\label{Fact1}
 {\bf L}(z)={\bf L}_+(z){\bf L}_-(z)\qquad \text{ with } \quad {\bf L}_{\pm}(z)=\text{diag}(\mathfrak{L}^{(1)}_{\pm}(z), \mathfrak{L}^{(2)}_{\pm}(z))\;,
\end{equation}
and $z\in{\mathcal{A}}$.
The factors $\mathfrak{L}^{(j)}_{\pm}(z)$, $j=1,2$ are defined via the Cauchy integral:
\begin{equation}\label{Fact2}
\mathfrak{L}^{(j)}_{\pm}(z)=\exp\left(\pm \frac{1}{2\pi {\rm i}}\oint_{\mathcal{C}}\frac{\log \mathfrak{L}_j(\xi)}{z-\xi}{\rm d} \xi\right),
\end{equation}
where $\mathcal{C}$ is any rectifiable, closed contour oriented counter-clockwise in the annulus ${\mathcal{A}}$ (see Section \ref{sec2p2} and \ref{sec2p3}). Note that ${\mathcal{A}}$ is also an annulus of analyticity of the kernel functions $\mathfrak{L}_j(z)$, $j=1,2$ where these functions also do not vanish. 
In addition, 
\begin{equation}\label{eqFj}
\mathrm{f}_-^{(j)}(z)=-\frac{z}{2(z-w_j)}\;,\qquad w_j=(-1)^{j} e^{{\rm i}k_{\tt x}}\;,\quad j=1,2,
\end{equation}
for $z\in{\mathcal{A}}.$

The scattered field can then be obtained in the form of a contour integral by inverting the transform \eqref{ZT}. Based on (\ref{u_upper}) and (\ref{SOL}), this leads to:
\begin{equation}\label{UxyR} {{\mathbf{u}}}^{\mathfrak{s}}_ {{\tt x}, {\tt y}}=\left\{\begin{array}{ll}
\displaystyle{\frac{1}{2\pi {\rm i}} \oint_{\mathcal{C}} {\bf P}({\tt y}) \mathbf{u}^{\rm F}_0 z^{{\tt x}-1} {\rm d} z} &\qquad \text{ for } {\tt y}\in{{\mathbb{Z}}_+}
\\ \\
\displaystyle{\frac{1}{2\pi {\rm i}} \oint_{\mathcal{C}} {\bf Q}(-{\tt y}) \mathbf{u}^{\rm F}_0 z^{{\tt x}-1} {\rm d} z } &\qquad \text{ for } {\tt y}\in{{\mathbb{Z}}_-}\;,
\end{array}\right.
\end{equation}
for ${\tt x}, {\tt y}\in \mathbb{Z}^2$, where 
\begin{equation}\label{sym2}
{\bf P}({\tt y})=[P_{ij}({\tt y})]_{i,j=1}^2={\bf B \rm}\Lambda^{{\tt y}} {\bf B}^{-1} \quad \text{ and } \quad {\bf Q}({\tt y})=[Q_{ij}({\tt y})]_{i,j=1}^2={{{\textbf J}}}{\bf B \rm}\Lambda^{{\tt y}} {\bf B}^{-1}{{\textbf J}} \;.
\end{equation}
The total field ${{\bf u}}_ {{\tt x}, {\tt y}}$ then follows by combining the above with (\ref{Us}) and the incident wave in (\ref{incscasum}). 
\begin{figure}[!ht]
\center{\includegraphics[width=.8\textwidth]{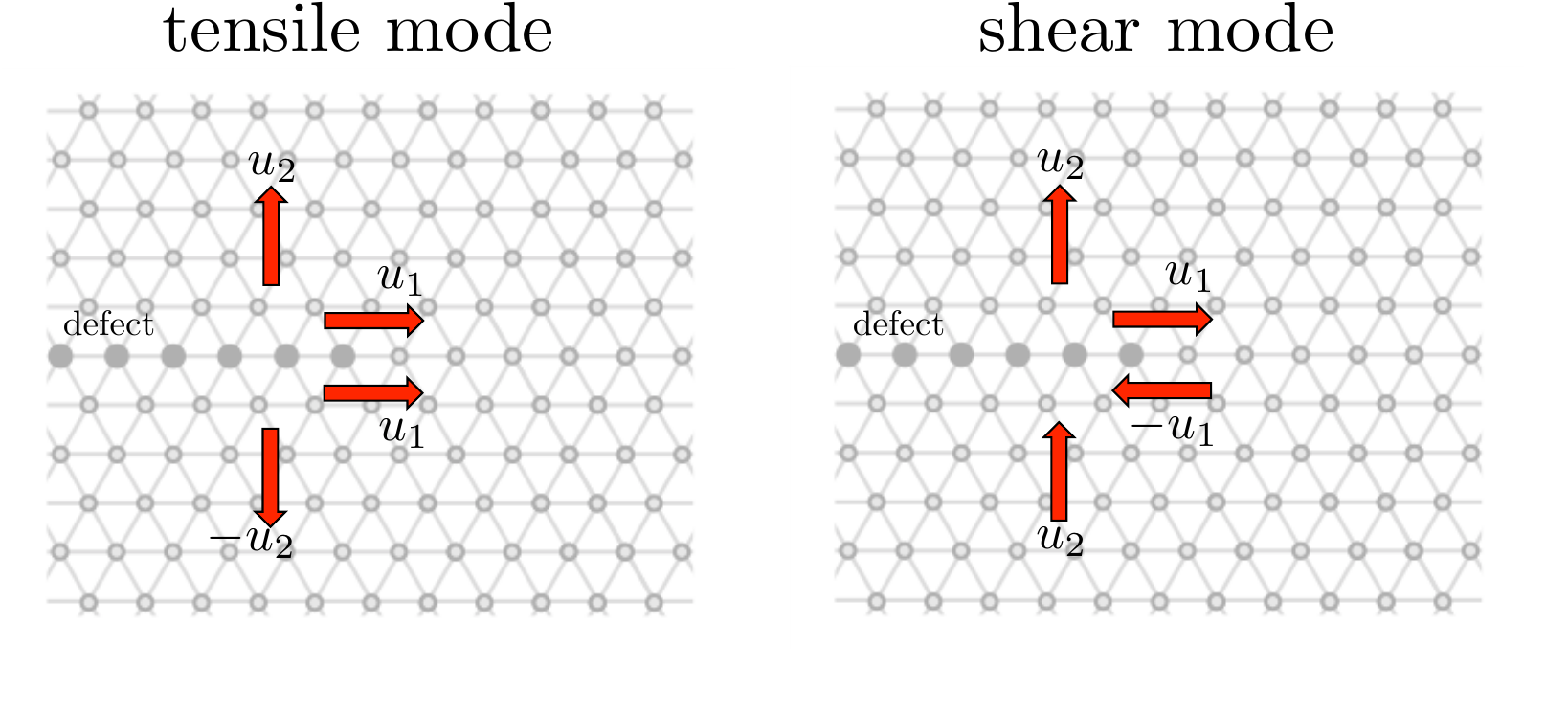}}
(a)~~~~~~~~~~~~~~~~~~~~~~~~~~~~~~~~~~~~~~~~~~~~~~~~~~~~~~~~~~~~~(b)
\caption{{The dynamic modes describing the far-field lattice behaviour during the interaction of an incoming wave with the inertial defect.} (a) The tensile mode associated with the symmetry class $\mathcal{T}$, where the horizontal (vertical) displacements, denoted by $u_1$ ($u_2$) are even (odd) functions with respect to the lattice row containing the defect. 
(b) The shear mode connected with the symmetry class $\mathcal{S}$, whose horizontal displacements $u_1$ ($u_2$) are odd (even) functions with respect to the row containing the defect. The third class of modes, or the mixed mode class $\mathcal{M}$ utilises a linear combination of modes in (a) and (b).}
\label{pic_modes}
\end{figure}
\section{Scattering response of the lattice}\label{sec4}

\subsection{Propagation of the bulk incident wave}
Recall that according to Section \ref{inc_w_sec}, the scattering of elastic plane waves is only possible in the frequency range $\omega <\sqrt{6}$, where the pass band for the bulk medium exists. Within this frequency regime, an incident wave can be defined as outlined in Section \ref{inc_w_sec}.
The frequency $\omega=\sqrt{6}$ is the cut-off frequency for the upper dispersion surface that is connected with pressure waves in the low frequency regime. Further, there is a cut-off frequency at $\omega=3\sqrt{2}/2$ for the lower dispersion surface, which is associated with shear waves at low-frequencies. Below this frequency, one can penetrate the inertial defect with a combination two different waveforms, whereas this is only possible with one waveform in the range $3\sqrt{2}/2<\omega<\sqrt{6}$.

Close to the specified cut-off frequencies within the bulk pass-band, we may observe dynamic anisotropy in the scattering response of the inertial defect. This is connected with the topology of the dispersion surfaces for the bulk lattice (see (\ref{incdisp})). Those surfaces exhibit non-circular slowness contours near the cut-off frequencies. These contours promote preferential directions of wave propagation in the lattice that appear in the scattered field. Some effects observed in this context, prevalent in the scattering process, include the appearance of star-shaped scattered wave forms, negative reflection and negative refraction. These effects are explored later in the illustrative examples.

\subsection{Structure of the scattered field far from the defect tip}
Far from the origin in the lattice, the contour integral representation of the scattered field (\ref{UxyR}) can be written as a linear combination of terms corresponding to the singular points $z\in \mathbb{C}$ of the integrand. Each of these singular points represents a
lattice wave induced by the inertial defect when interacting with the incident wave.
{Further, it is the leading order asymptotes of $\mathbf{u}^{\rm F}_0$ near these singular points that dictate the form of the excited waves. }

{Thus, in the far-field we write the scattered field as:
\begin{equation}\label{mode_rep}
{\mathbf{u}}^{\mathfrak{s}}_ {{\tt x}, {\tt y}}=\sum_{j=1,2} {\mathbf{u}}^{\text{bulk}}_ {{\tt x}, {\tt y}}(w_j)+\sum_{z_0\in {\mathcal{Z}}}{\mathbf{u}}^{\text{behind}}_{{\tt x}, {\tt y}}(z_0)+\sum_{z_b\in {\mathcal{B}}}{\mathbf{u}}^{\text{ahead}}_ {{\tt x}, {\tt y}}(z_b)+{\mathbf{u}}^{\text{rem}}_ {{\tt x}, {\tt y}}.
\end{equation}
as $|{\bf x}| \to\infty$, ${\bf x}=(\tilde{x}_1, \tilde{x}_2)^{\rm T}$ (see also Appendix \ref{App_mode_rep} for the derivation of the preceding formula). 
}

In \eqref{mode_rep},
\begin{enumerate}
\item ${\mathbf{u}}^{\text{bulk}}_ {{\tt x}, {\tt y}}$ represents the reflection and transmission of the incident wave through its interaction with the inertial defect. These terms are associated with the simple poles $w_j$, $j=1,2,$ of ${\mathbf{u}}^{\rm F}_0$ in (\ref{SOL}).

\item ${\mathbf{u}}^{\text{behind}}_{{\tt x}, {\tt y}}(z_0)$,
with $z_0 \in \mathcal{Z}$ being simple poles of (\ref{SOL}) and
\begin{equation}\label{z0set}
 \mathcal{Z}:=\{z\in \mathbb{C}, |z|>1: {\mathfrak{L}_j}(z)=0, j=1,2\},
\end{equation}
represents localised wave modes produced by the dynamic response of the inertial defect.

\item ${\mathbf{u}}^{\text{ahead}}_ {{\tt x}, {\tt y}}(z_b)$, with $z_b \in \mathcal{B}$ being singular branch points of (\ref{SOL}) and where
\begin{equation}\label{zbset}
\mathcal{B}:=\{z\in \mathbb{C}, |z|<1: ({\mathfrak{L}_j}(z))^{-1}=0, j=1,2\},
\end{equation}
represents the waves that radiate away from the inertial defect and whose amplitude decays into the bulk lattice. These waves are connected with mechanical oscillations produced ahead by inertial defect during the scattering process.
\item {${\mathbf{u}}^{\text{rem}}_ {{\tt x}, {\tt y}}$ involves the contour integral of a complex function over the unit disk. The associated integrand has  no singular points or zeros in $\mathcal{C}$ but possesses saddle points along this contour that provide a non-negligible contributions to the far-field lattice behave. Those contributions engage  with terms ${\mathbf{u}}^{\text{ahead}}_ {{\tt x}, {\tt y}}(z_b)$ to ensure the combined effect tends to zero as $O(1/\sqrt{|{\bf x}|})$ (see Section 3 and Appendix C of \cite{Bls1}) as $|{\bf x}| \to\infty$, ${\bf x}=(\tilde{x}_1, \tilde{x}_2)^{\rm T}$.}
\end{enumerate}
{In the above, if ${\mathcal{B}}=\varnothing$ or ${\mathcal{Z}}=\varnothing$ then the corresponding sums in (\ref{mode_rep}) are taken as zero. Further, those points in ${\mathcal{B}}$ and ${\mathcal{Z}}$ situated a finite distance from the unit disk in $\mathbb{C}$ are linked to modes that decay exponentially away from the defect tip in their region of support. 
On the other hand, those points in ${\mathcal{B}}$ and ${\mathcal{Z}}$ located in the immediate vicinity of the unit disk in $\mathbb{C}$ correspond to wave modes with a different behaviour. {For those in ${\mathcal{B}}$, they have a square root type decay into the bulk accompanied by oscillations arriving from the combination of the last two terms (\ref{mode_rep}).} For points in $\mathcal{Z}$, sinusoidal oscillations are present in the direction parallel to the defect and these modes decay exponentially in direction perpendicular to the defect. 
Below, we limit the analysis of the dispersive nature of the system to the study of these wave modes corresponding to complex $z$ in the vicinity of the unit disk in $\mathbb{C}$.}

\subsection{Symmetry properties of the lattice wave modes}
\label{symmsec}
The matrices in (\ref{sym2}) are important in characterising symmetry properties about $n_2=0$ of the wavemodes ${\mathbf{u}}^{\text{behind}}_{{\tt x}, {\tt y}}$ and ${\mathbf{u}}^{\text{ahead}}_{{\tt x}, {\tt y}}$ (see (\ref{mode_rep})) appearing in the scattering process. Indeed, we note that the entries of the matrices in (\ref{sym2}) satisfy
\begin{equation}\label{sym1}
 P_{ij}({\tt y})=(-1)^{i+j}Q_{ij}({\tt y}) \quad \text{ for } \quad {\tt y}\in{{\mathbb{Z}}_+} \text{ and }\quad 1\le i, j \le 2\;.
 \end{equation}
 Thus if, to leading order, the asymptote of the vector function $\mathbf{u}^{\rm F}_0$ is proportional to {${\bf e}_1=(1,0)^{\rm T}$ or ${\bf e}_2=(0,1)^{\rm T}$}, the above fact together with (\ref{UxyR}) implies the lattice wave, respectively, belongs to either: 
\begin{enumerate}[$\bullet$]
\item a {\bf symmetry class $\mathcal{T}$} representing dynamic tensile modes (see Figure \ref{pic_modes}(a)) whose first and second displacement components are even and odd functions of $n_2$, respectively, about the lattice row at $n_2=0$. Mathematically, these lattice waves appear due to the singular points exclusive to ${L}_+^{(1)}(z)$ and zeros of ${L}_-^{(1)}(z)$ in (\ref{SOL}) that dictate the response ahead or behind the edge of the inertial defect, respectively.

\item a {\bf symmetry class $\mathcal{S}$}, whose modes are dynamic shear modes as illustrated in Figure \ref{pic_modes}(b). The shear modes involves displacements whose lateral and transverse components are governed by odd and even functions of $n_2$, respectively, about $n_2=0$. The singular points exclusive to ${L}_+^{(2)}(z)$ or the zeros of ${L}_-^{(2)}(z)$ in (\ref{SOL}) promote the appearance of waves in this class.
\end{enumerate}
It may also happen that the leading order asymptote of $\mathbf{u}^{\rm F}_0$ near its singular point is a vector with non-zero entries. In this case, we also define the:

\begin{enumerate}[$\bullet$]
\item {\bf mixed mode class $\mathcal{M}$}, formed from linear combinations of wave modes from the symmetry class $\mathcal{T}$ and $\mathcal{S}$. These lattice waves are linked to points $z\in \mathbb{C}$ where ${L}_+^{(j)}(z)$, $j=1,2$ are both singular or ${L}_-^{(j)}(z)$, $j=1,2$  are both zero in (\ref{SOL}). Later, we see that some of these waves are connected with dispersion curves for the lattice chain located at ${\tt x}\ge 0$, which appear inside narrow non-zero frequency pass bands. Further, these waves also occur in connection with crossings or exceptional points of the dispersion curves concerning the dynamic response of the inertial defect and the bulk lattice.
\end{enumerate}

{Further, if a wave mode in the structure is defined by a $z$ value and $\lambda_{1,2}(z)$ (see (\ref{lam_sol})) located in the vicinity of the unit-circle in the complex plane, the diagonal and off-diagonal elements of ${\bf P}({\tt y})$ are real and purely imaginary, respectively. In addition, in this case ${\bf P}({\tt y})=\overline{{\bf Q}({\tt y})}$, with the over line denoting the complex conjugate. Note that ${\bf u}_0^{\rm F}$ couples the columns of these matrices in (\ref{UxyR}) and this can create disparate behaviours in the structure for ${\tt y}\ge 0$ and ${\tt y}<0$ depending on the strength of this coupling. This is examined in the examples that follow.} 

\vspace{0.1in}{\bf \emph{Reflection and transmission of the incident wave by the defect. }}Finally, we note the Wiener-Hopf solution (\ref{SOL}) also contains simple poles belonging to ${\cal F}_{-}^{(j)}$, $j=1,2$. They give rise to ${\mathbf{u}}^{\text{bulk}}_ {{\tt x}, {\tt y}}(w_j)$, $j=1,2$, in (\ref{mode_rep}) that represent reflected and transmitted waveforms appearing when the incident wave interacts with structured inertial defect. These non-symmetric wave modes propagate away from the inertial defect to the remote regions of the lattice with a constant amplitude. These features may also be enhanced by wave modes appearing when the inertial defect oscillates, which only occurs when $\mathfrak{L}_-^{(j)}(z)$, $j=1,2,$ have zeros in the immediate vicinity of the unit circle in $\mathbb{C}$.

\begin{figure}[!ht]
\center{\includegraphics[width=0.6\textwidth]{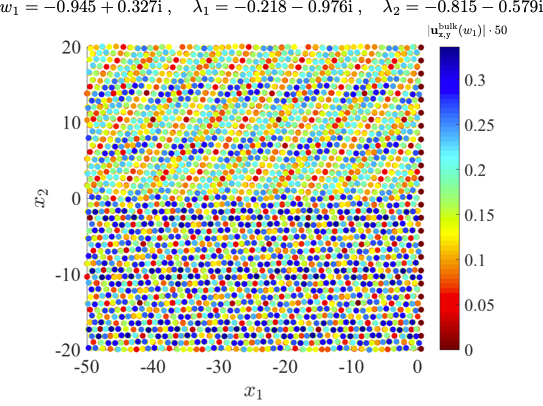}}

\caption{{Mode produced ${\bf u}^{\rm bulk}_{\tt x, y}(w_1)$ behind a heavy defect ($\beta=1$) interacting with an incoming wave with the frequency $\omega=1$, amplitude $0.01$ and angle of incidence $47^\circ$ measured anticlockwise from the positive $x_1$-axis (see Section \ref{sec2} for the definition of $x_j$, $j=1,2$). This computation is based on (\ref{ubulk}) in Appendix \ref{appbulk}.}
}
\label{pic_mode_w1}
\end{figure}

{Figure \ref{pic_mode_w1} shows an example of a wave mode ${\bf u}^{\rm bulk}_{\tt x, y}(w_1)$ produced a heavy defect interacting with an incident wave as specified in the caption. For the sake of brevity, we do not report the mode ${\bf u}^{\rm bulk}_{\tt x, y}(w_2)$, which produces a similar response in the considered system. In Figure \ref{pic_mode_w1}, it can be seen the field contributing to the reflected component of the incident wave (in ${\tt y}\ge 0$) has an apparent wavefront perpendicular to  $135^{\circ}$ (measured anticlockwise from the positive $x_1$-axis). The transmitted component located in the lower half-plane possesses a more complex behaviour and this wave appears to have been refracted upon penetrating the inertial defect. Mathematically, the contrast between the behaviour of the structure in ${\tt y}\ge 0$ and ${\tt y}<0$ is due to the aforementioned coupling of the columns of ${\bf P}({\tt y})$ and ${\bf Q}({\tt y})$. The form of the modes ${\bf u}^{\rm bulk}_{\tt x, y}(w_j)$, $j=1,2,$ are given in Appendix \ref{appbulk}.}

 The above lattice waves are discussed in the next sections in conjunction with the dispersive properties of the medium and computations concerning the lattice scattering response.

\subsection{Dynamic response ahead of the inertial defect}\label{scat_resp}

\subsubsection{Bulk lattice waves generated by the inertial defect}\label{scat_bulk}
 Waves modes ${\mathbf{u}}^{\text{ahead}}_ {{\tt x}, {\tt y}}$ in (\ref{mode_rep}) appear as a result of singular branch points of ${\mathbf{u}}_0^{\rm F}$.
 For $z\to z_b$, with $z_b\in \mathcal{B}$, 
\begin{equation}\label{asympbranch}
{\mathbf{u}}^{\rm F}_0\sim {\bf M}(z_b)\Big({1-\frac{z_b}{z}}\Big)^{-1/2}\;,
\end{equation}
where 
\begin{equation}\label{Mzb}
{\bf M}(z_b)={\bf V}(z_b)\sum_{j=1}^2 [{\bf L}_+(w_j)]^{-1}{\bf a}^{\mathfrak{i}}\mathrm{f}^{(j)}_-(z_b),
\end{equation}
\[{\bf V}(z_b)=\left\{\begin{array}{ll}
-{L}^{(1, *)}_+(z_b)({\bf e}_1 \otimes {\bf e}_1) \;, & \quad \text{for symmetry $\mathcal{T}$,}\\[1ex]
-{L}^{(2, *)}_+(z_b)({\bf e}_2 \otimes {\bf e}_2)\;, & \quad \text{\rm for symmetry $\mathcal{S}$,}\\[1ex]
-{\bf L}^{(*)}_+(z_b)\;, & \quad \text{\rm for mixed modes in $\mathcal{M}$,}
\end{array}\right.\]
with 
\[{L}^{(j, *)}_+(p):=\lim_{z\to p}\sqrt{1-\frac{p}{z}}\, {L}^{(j)}_+(p), \quad j=1,2,\; \quad \text{ and }\quad {\bf L}^{*}_+(p):=\lim_{z\to p}\sqrt{1-\frac{p}{z}}\, {\bf L}_+(p)\;.\]
{The notation ${\bf a}\otimes{\bf b}=[a_i b_j]_{i,j=1}^2$ represents the tensor product of two vectors ${\bf a}, {\bf b}\in{\mathbb{R}}^2$. Details of the derivations of the asymptotes in (\ref{asympbranch}) are given in Appendix \ref{secF1}. Then we combine (\ref{asympbranch}) with (\ref{UxyR}) and use the result
\begin{equation}\label{ointA}
\oint_{{\mathcal{C}}}z^{{\tt x}-1}\Big(1-\frac{z_b}{z}\Big)^{-1/2}\, {\rm d} z=2{\rm i}z_b^{\tt x}\,{\rm B}({\tt x}+1/2, 1/2)H({\tt x})\;,
\end{equation}
proved in Appendix \ref{OintAsec}. In doing so, we obtain
 the asymptotes (\ref{asympbranch}) then give} contributions to the inverse transform in (\ref{UxyR}) that give the provide of ${\mathbf{u}}^{\text{ahead}}_ {{\tt x}, {\tt y}}$ as:
 \begin{equation}
{\mathbf{u}}^{\text{ahead}}_ {{\tt x}, {\tt y}}(z_b)={\bf N}_{{\tt y}}(z_b){\rm B}(1/2, {\tt x}+1/2)z_b^{{\tt x}}H({\tt x})
\label{mode_rep_ahead}
\end{equation}
with 
\[{\rm B}(p, q) =\int_0^1 t^{p-1} (1-t)^{q-1}\, {\rm d}t\;,\]
being the Beta function and
\begin{equation}\label{Ny}{\bf N}_{{\tt y}}(z_b)=\left\{\begin{array}{ll}
\displaystyle{ \pi^{-1}{\bf P}({\tt y}){\bf M}(z_b)}, &\qquad \text{ for } {\tt y}\in{{\mathbb{Z}}_+},
\\[1ex]
\displaystyle{\pi^{-1}{\bf Q}(-{\tt y}){\bf M}(z_b)}, &\qquad \text{ for } {\tt y}\in{{\mathbb{Z}}_-}\;.
\end{array}\right.
\end{equation}
Here, ${\bf N}_{{\tt y}}$ is responsible for supplying the mode with the required symmetry properties in accordance with the symmetry class that describes the mode. This depends on the connection between $z_b$ and the dispersive properties of the bulk lattice which is discussed in the next section.

\subsubsection{{Numerical examples of the modes} \texorpdfstring{${\bf u}^{\rm ahead}_{\tt x, \tt y}$}{}}
\begin{figure}[!ht]
\center{\includegraphics[width=1\textwidth]{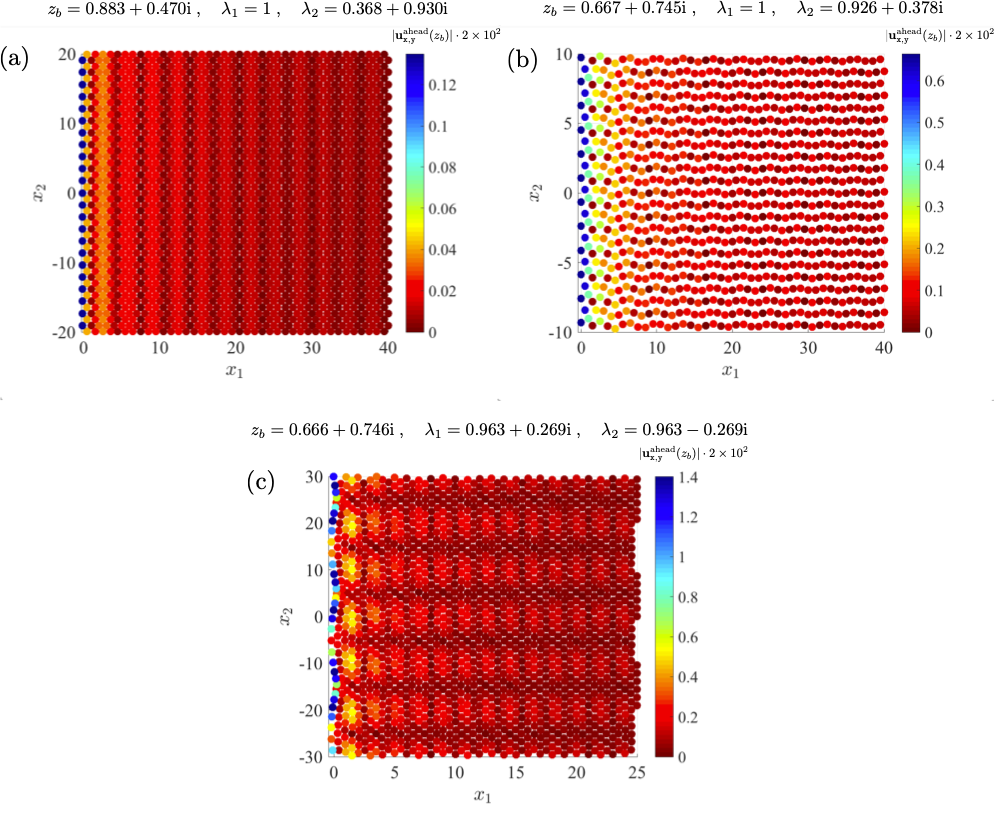}}

\caption{Modes produced ahead of a heavy defect ($\beta=1$) interacting with an incoming wave with the frequency $\omega=1$, amplitude $0.01$ and angle of incidence $47^\circ$ measured anticlockwise from the positive $x_1$-axis (see Section \ref{sec2} for the definition of $x_j$, $j=1,2$). (a) A symmetric mode. (b) An antisymmetric mode. (c) A mixed mode. The computations shown here are based on (\ref{mode_rep_ahead}). 
{Note that in these cases $|\lambda_1|=|\lambda_2|=1$ and therefore the modes do not decay along $x_2$ direction; see item 4 following \eqref{mode_rep} that explains overall decay of the diffracted field away from the defect-tip.}
}
\label{pic_mode_ahead}
\end{figure}

{As expected, the above representation reveals the scattered bulk wave modes produced by the inertial defect exhibit a square-root decay in the {$x_1$-direction}, provided by the Beta function, in far-field of the lattice. This behaviour is also accompanied by oscillations if $|z_b|=1$, whereas these wave modes decay exponentially if $|z_b|<1$, $z_b\in \mathcal{B}$. {Note that the representations (\ref{mode_rep_ahead}) combine with  ${\mathbf{u}}^{\text{rem}}_ {{\tt x}, {\tt y}}$ in (\ref{mode_rep}) to provide a response that decays in the far-field of the lattice.}

Figure \ref{pic_mode_ahead}(a) shows a symmetric mode generated by the lattice in the case when a wave incident on a heavy defect at an angle of 47$^{\circ}$, measured anticlockwise and with frequency $\omega=1$ from the positive $x_1$-axis, and having amplitude $0.01$. Here, this is a special case when ${\bf N}_{\tt y}$ is proportional to $\mathbf{e}_1$ and invariant with respect to ${\tt y}$ (${\bf N}_{\tt y}=(-6.748-3.717 {\rm i}, 0)^{\rm T} \times 10^{-4}$). In this case, the corresponding mode ${\bf u}^{\rm ahead}_{\tt x, \tt y}$ is represented as the translation of all lattice nodes ahead of the defect tip parallel to the $x_1$-axis. }

Figure \ref{pic_mode_ahead}(b) shows the anti-symmetric mode involved in the same scattering process where ${\bf N}_{\tt y}$ is now proportional to $\mathbf{e}_2$, as well as being independent of ${\tt y}$ (here ${\bf N}_{\tt y}=(0, -3.319+5.361 {\rm i}, 0)^{\rm T} \times 10^{-3}$). Hence, the dynamic contribution of the mode to the system's response is to supply a translation to all nodes parallel to the $x_2$-axis ahead of the defect, as illustrated in Figure \ref{pic_mode_ahead}(b).

Finally, Figure \ref{pic_mode_ahead}(c) demonstrates the behaviour of ${\bf u}^{\rm ahead}_{\tt x, \tt y}$ from the mixed mode class $\mathcal{M}$. The resulting ${\bf N}_{\tt y}$ now produces a translation to the nodes in both the $x_1-$ and $x_2-$directions and depends on ${\tt y}$. Correspondingly, Figure \ref{pic_mode_ahead}(c) shows the classical symmetry expected about $n_2=0$ from the modes in the sets $\mathcal{T}$ and $\mathcal{M}$ is no longer present. This lack of this symmetry is most visible near $x_1=0$. {For this mode, as described in Section \ref{symmsec}, we have ${\bf P}({\tt y})=\overline{\bf Q}({\tt y})$, where ${\bf P}$ has imaginary off-diagonal components and real diagonal components. If ${\bf M}(z_b)$ in (\ref{Mzb}) has a dominant component, the coupling between the columns of ${\bf P}({\tt y})$ (that plays a role in the far-field behaviour of the system, see (\ref{UxyR})) in the corresponding mode is weak. Consequently, the behaviour of the lattice for ${\tt y}\ge 0$ and ${\tt y}<0$ appears to be very similar. This is the case in Figure \ref{pic_mode_w1}(c), where ${\bf M}(z_b)=(2+2{\rm i}, 70+50{\rm i})^{\rm T}\times 10^{-4}$. This is contrary to the case of Figure \ref{pic_mode_w1}, where the coupling of those columns of ${\bf P}({\tt y})$ and ${\bf Q}({\tt y})$ is strong and one can observe a completely different behaviour between the upper and lower half-spaces (see also Appendix \ref{appbulk} for further details).}

Lastly, the modes presented here in Figures \ref{pic_mode_w1} and \ref{pic_mode_ahead} are a sub-collection of those that are important in characterising the remote lattice response ahead of defect. The full scattering response of the system in this particular case is shown in Figure \ref{pic_beta_1} (see Section \ref{sec5}).

In the next section, we discuss the dispersion curves that define the waves ${\mathbf{u}}^{\text{ahead}}_ {{\tt x}, {\tt y}}$ possessing the oscillatory behaviour, i.e. when $|z_b|=1$.

\subsubsection{Dispersive properties ahead of the inertial defect}
\label{sec_disp_bulk}
The contributions ${\mathbf{u}}^{\text{ahead}}_ {{\tt x}, {\tt y}}$ in (\ref{mode_rep}) and (\ref{mode_rep_ahead}) that embed the oscillatory behaviour are linked to dispersion curves that define wave modes ahead of the inertial defect along $n_2=0$. The pass band for these mechanical oscillations involves dispersion curves that promote modes possessing symmetry $\mathcal{T}$, symmetry $\mathcal{S}$ and mixed mode responses in $\mathcal{M}$ in the structured medium. 

Dispersion curves for the chain (inertial defect) located along $n_1 \in{{\mathbb{Z}}_+}, n_2=0$ ($n_1\in{{\mathbb{Z}}_-}, n_2=0$) are readily analysed by associating the complex variable $z$ with the normalised wavenumber $k_1$ through $k_1=e^{{\rm i}k_1/2}$
(see also $\eqref{k1k2}{}_1$, $(\ref{dim5})$),
where $z$ is taken as any frequency dependent singular branch point (or zero) of the functions $\mathfrak{L}_j(z)$, $j=1,2$. 

\begin{figure}[!ht]
\center{\includegraphics[width=0.7\textwidth]{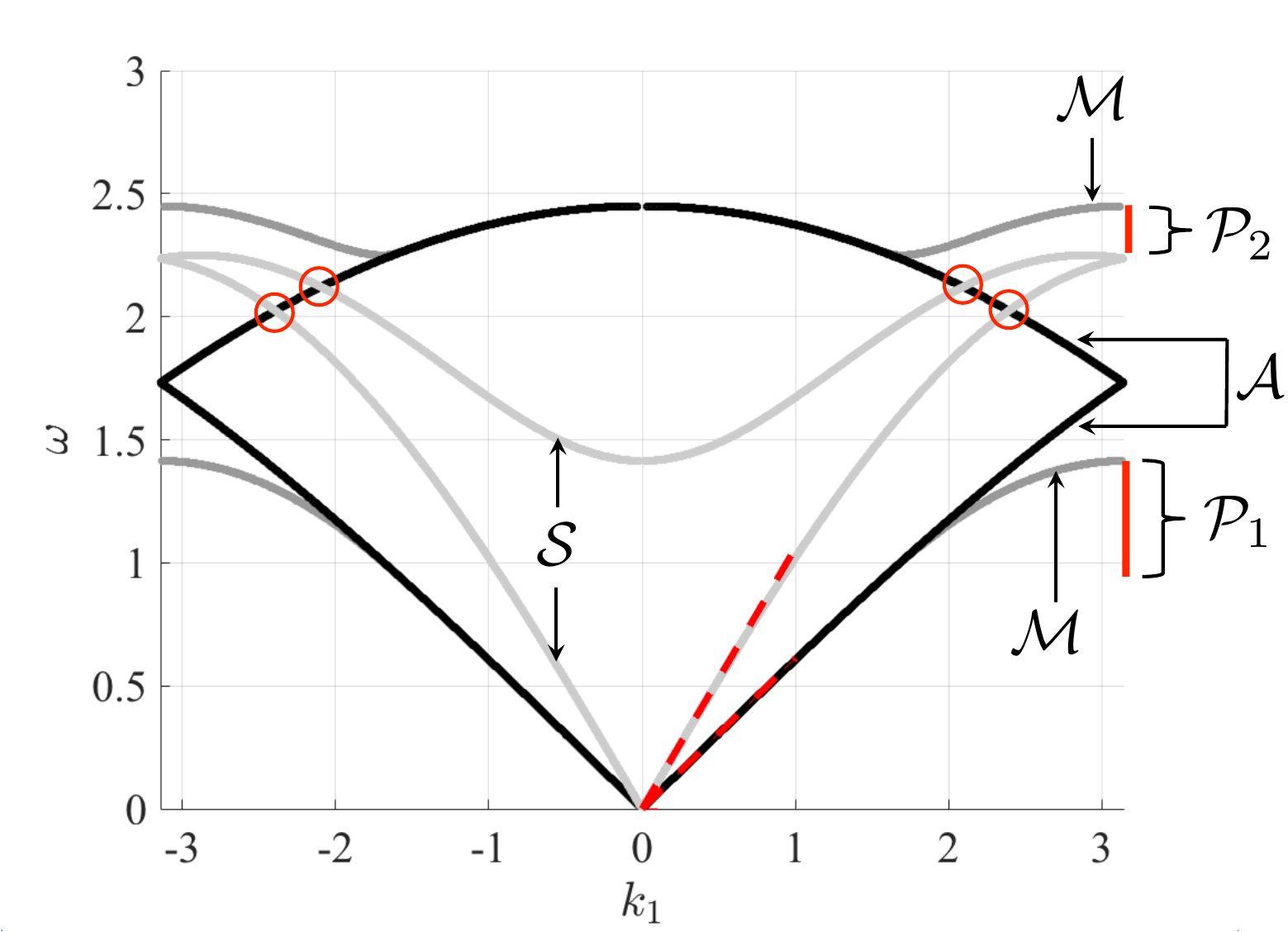}}
\caption{Dispersion curves for the bulk medium external to the dynamic inclusion based on identifying the singular branch points of $\mathfrak{L}_j(z), j=1,2$ (see (\ref{L1L2})). The curves are based on the results in (\ref{eqrootconj}), (\ref{z3pmA}), (\ref{z1pm}) and (\ref{z2pm}) in Appendices \ref{Singular1} and \ref{Singular2}. Dashed red lines are given by $\omega=\sqrt{8/3}k_1$ and $\omega= \sqrt{9/8}k_1$. Red bands to the right of the diagram, labelled $\mathcal{P}_1$ and $\mathcal{P}_2$, are the passbands for the dispersion curves for the mixed mode class $\mathcal{M}$. Red circles correspond to points where dispersion curves for the symmetry classes ${\mathcal{T}}$ and $\mathcal{S}$ cross.}
\label{pic_dd_bulk}
\end{figure}

At this juncture, we concentrate the analysis on real values of $k_1$, i.e. those zeros and singular branch points of ${\mathbf{u}}_0^{\rm F}$ in (\ref{SOL}) located along the unit circle in the complex plane. As an example of this analysis, in Figure \ref{pic_dd_bulk} we show the dispersion curves connected with wave motion ahead of the inertial defect along $n_2=0$ in the lattice.

The curves of Figure \ref{pic_dd_bulk} are associated with three types of dynamic phenomena or bulk lattice waves created in the scattering process induced by the inertial defect. {These lattice waves represent:
\begin{enumerate}[$\bullet$]
\item symmetry $\mathcal{T}$ (light grey curves in Figure \ref{pic_dd_bulk}, labelled as ``$\mathcal{T}$" and computed from (\ref{z3pmA}) of Appendix \ref{Singular2}). Along these curves, one can verify ${\bf P}({\tt y})$ is upper triangular and the corresponding modes ${\mathbf{u}}^{\text{ahead}}_ {{\tt x}, {\tt y}}(z_b)$ particular symmetric modes involving only  translations of the lattice nodes in the $x_1$-direction. 
\item symmetry $\mathcal{S}$ (black curves, indicated by ``$\mathcal{S}$" and based on (\ref{z1pm}) and (\ref{z2pm}) in Appendix \ref{Singular2}). Here, it is possible to show ${\bf P}({\tt y})$ is lower triangular and the associated modes ${\mathbf{u}}^{\text{ahead}}_ {{\tt x}, {\tt y}}(z_b)$ particular anti-symmetric modes having only translations in the $x_2$-direction.
\item  {mixed modes} (dark grey curves marked with ``$\mathcal{M}$" and identified from (\ref{eqrootconj}) of Appendix \ref{Singular1}). 
\end{enumerate}}
Mathematically, these lattice waves are connected with singular branch points of $\mathfrak{L}^{(j)}_-(z)$, $j=1, 2$, (see Section \ref{symmsec}) 
located in the interior of the unit disk of the complex plane. These are routinely found by analysing the singular points of Wiener-Hopf kernel.

{\bf \emph{Acoustic dispersion curves:}} Figure \ref{pic_dd_bulk} shows that one of each of the dispersion curves for symmetry $\mathcal{S}$ and symmetry $\mathcal{T}$ wave motion are of the acoustic type, i.e. they intersect with the origin of the $(k, \omega)$-space. In the symmetry class $\mathcal{S}$, the low-frequency wave motion associated with the corresponding acoustic curve is connected shear waves, whereas for the symmetry class $\mathcal{T}$ the low-frequency eigenfields are pressure waves. Further, these acoustic curves correspond to the classical linear dispersion relations for pressure and shear waves in a continuous elastic medium, i.e. $k_1\to 0$ (see Section \ref{inc_w_sec}). 

In addition, the acoustic dispersion curves connected with the classical continuum model show there are low-frequencies where mode conversion in the lattice can occur during the scattering process. In particular, for a given low-frequency, it can happen that a shear wave incident on the inertial defect can generate pressure waves when scattered and vice versa.

Note that the acoustic branches also coincide with dispersion curves obtained from (\ref{incdisp}), that define the incident field when $k_2=0$, with the only difference that the lattice response associated with both pairs of curves is very different. In this context, (\ref{incdisp}) is used to define a constant amplitude wave incident on the defect and the {scattered} wave {geometrically} reflected {above the defect and} transmitted {below} the inertial defect. The latter are defined by the simple poles in ${{\rm f}_-}$ in the Wiener-Hopf solution (see (\ref{SOL}) and (\ref{eqFj})). 
On the other hand, these acoustic branches arise from the Wiener-Hopf analysis correspond to the wave modes induced by the inertial defect in the scattering problem. {Those modes decay in the far field of the lattice when combined with ${\mathbf{u}}^{\text{rem}}_ {{\tt x}, {\tt y}}$ in (\ref{mode_rep}) (see Section \ref{scat_resp})}.

{{\bf \emph{Optical dispersion curves:}} The remaining dispersion curves in Figure \ref{pic_dd_bulk} associated with symmetry $\mathcal{S}$ and symmetry $\mathcal{T}$ wave motion are akin to optical dispersion relations, i.e. they coincide with a non-zero frequency at $k_1=0$. Along these curves, $\lambda_1=-1$ or $\lambda_2=-1$. Hence, the lattice wave motion oscillates with respect to varying $n_2$ in the lattice. Compare this with the case of the acoustic branches, where along these curves these exponents are equal to unity and hence the analogous behaviour in lattice displacements is absent there. } 

The last collection of optical curves produce modes from the class $\mathcal{M}$.
Those curves only cover narrow (disjoint) intervals of the radian frequency e.g. for $0.94<\omega<\sqrt{2}$ and $2.25<\omega<\sqrt{6}$ (labelled $\mathcal{P}_1$ and $\mathcal{P}_2$, respectively, in Figure \ref{pic_dd_bulk}). Inside these intervals, the structured medium's response will exhibit dynamic anisotropy. The associated dispersion curves also exist in narrow intervals of the wavenumber, which is an unusual topological feature of these curves that has been previously identified in models of fracture in complex lattice systems \cite{Piccolroazetal}. In addition, we observe that in the band $\mathcal{P}_2$, there are no dispersion curves corresponding to symmetry class $\mathcal{T}$ wavemodes and only modes within the classes $\mathcal{M}$ and $\mathcal{S}$ are excited in this regime. Responses corresponding to scattering process for incident waves at a frequency located within $\mathcal{P}_2$ are reported in Figure \ref{pic_beta_m0p5}.
Further, Figure \ref{pic_dd_bulk} shows that between the bands $\mathcal{P}_1$ and $\mathcal{P}_2$, there exist exceptional points where the curves for the symmetry class $\mathcal{T}$ and $\mathcal{S}$ cross. At these frequencies, additional mixed modes can be excited in the bulk lattice. 

We note the optical branches identified here are not found in the associated low frequency-long wavelength model, and appear as a result of the lattice periodicity.

\subsection{Localised modes for the inertial defect}

\subsubsection{Wave modes for the inertial defect}
Wave modes ${\mathbf{u}}^{\text{behind}}_ {{\tt x}, {\tt y}}$ in (\ref{mode_rep}) are associated with dynamic response of the defect. Vibration modes of this structured inclusion correspond to simple poles $z_0\in \mathcal{Z}$ of ${\mathbf{u}}_0^{\rm F}$ outside the unit disk in the complex plane.
Indeed, when $z\to z_0$, the solution (\ref{SOL}) to the Wiener-Hopf equation admits the asymptote:
\begin{equation}\label{UxyRsimplpole}
{\mathbf{u}}^{\rm F}_0\sim {\bf D}(z_0)\Big(1-\frac{z}{z_0}\Big)^{-1} 
\end{equation}
with 
\[{\bf D}(z_0)= {\bf F}(z_0)
\sum_{j=1}^2 [{\bf L}_+(w_j)]^{-1}{\bf a}^{\mathfrak{i}}\mathrm{f}^{(j)}_-(z_0)\;,\]
and
\[{\bf F}(z_0)=\left\{\begin{array}{ll}
({L}^{(1, *)}_-(z_0))^{-1}({\bf e}_1 \otimes {\bf e}_1) \;, & \quad \text{for symmetry $\mathcal{T}$,}\\[1ex]
({L}^{(2, *)}_-(z_0))^{-1}({\bf e}_2 \otimes {\bf e}_2)\;, & \quad \text{for symmetry $\mathcal{S}$,}
\\[1ex]
({\bf L}^{(*)}_-(z_0))^{-1} \;, & \quad \text{for mixed modes},
\end{array}\right.\]
\[{L}^{(j, *)}_-(p)=\lim_{z\to p}\Big(1-\frac{z}{p}\Big)^{-1}\, {L}^{(j)}_-(p), \quad j=1,2 \quad \text{ and } \quad {\bf L}^{(*)}_-(p)=\lim_{z\to p}\Big(1-\frac{z}{p}\Big)^{-1}\, {\bf L}_-(p)\;.\]
When $|z_0|=1$, the symmetry $\mathcal{T}$ modes are associated with longitudinal wave motion of the particles in the inertial defect and this produces tensile-type responses in the lattice (see Figure \ref{pic_modes}(a)) On the other hand, transverse dynamic motion of the particles in inertial defect occur when the wave mode belongs to the symmetry class $\mathcal{S}$ and the associated bulk lattice response involves a shear dynamic mode (see Figure \ref{pic_modes}(b)). Finally, the mixed modes in the class $\mathcal{M}$ and produced by the vibrating defect only occur at exceptional points in the $(k, \omega)$-space, where the dispersion curves for longitudinal and transverse inertial defect wave motion intersect. 

The contributions of the simple poles to the integrals in (\ref{UxyR}) are readily calculated with the residue theorem. {To evaluate those contributions, we use the combination of (\ref{UxyRsimplpole}) together with (\ref{UxyR}) and the result 
\begin{equation}\label{ointsimppolesA}
\oint_{\mathcal{C}}\frac{z^{{\tt x}-1}}{1-z z_0^{-1}}{\rm d} z=2\pi {\rm i}z_0^{\tt x}H(-{\tt x})\;,
\end{equation}
proved in Appendix \ref{secF3}.
Those simple poles then} provide the localised defect modes ${\mathbf{u}}^{\text{behind}}_{{\tt x}, {\tt y}}$ in (\ref{mode_rep}) in the form:
\begin{equation}\label{ubehind}
{\mathbf{u}}^{\text{behind}}_{{\tt x}, {\tt y}}(z_0)={\bf S}_{{\tt y}}(z_0)z_0^{{\tt x}}H(-{\tt x})
\end{equation}
where
\begin{equation} \label{Sy}{\bf S}_{ {\tt y}}(z_0)=\left\{\begin{array}{ll}
\displaystyle{ {\bf P}({\tt y}){\bf D}(z_0)}, &\qquad \text{ for } {\tt y}\in{{\mathbb{Z}}_+},
\\[1ex]
\displaystyle{{\bf Q}(-{\tt y}){\bf D}(z_0)}, &\qquad \text{ for } {\tt y}\in{{\mathbb{Z}}_-}\;.
\end{array}\right.
\end{equation} 
In analogy with the modes generated ahead of the defect (discussed in Section \ref{scat_bulk}), the vector function ${\bf S}_{ {\tt y}}$ provides the symmetry properties of the modes ${\mathbf{u}}^{\text{behind}}_{{\tt x}, {\tt y}}$, in correspondence with the class $\mathcal{T}$, $\mathcal{S}$ or $\mathcal{M}$, depending on $z_0$. 

Every mode ${\mathbf{u}}^{\text{behind}}_{{\tt x}, {\tt y}}$ displays exponential localisation in the direction transverse to the defect. The strength of the localisation is dictated by the matrix $\Lambda$ (see (\ref{sym2}) and (\ref{lam_sol})).
Further, if $|z_0|>1$, the lattice wave ${\mathbf{u}}^{\text{behind}}_{{\tt x}, {\tt y}}(z_0)$ decays exponentially along the defect from its tip. When $|z_0|=1$, ${\mathbf{u}}^{\text{behind}}_{{\tt x}, {\tt y}}(z_0)$ represents a wave propagating along the inertial defect. If the frequency corresponding to the incident wave and generating such a mode coincides with the bulk lattice pass band, the vibration of the defect that may provide some enhancement to the reflected and diffracted wave components of the incident field.

\subsubsection{{Numerical examples of the modes ${\bf u}_{{\tt x},{\tt y}}^{\rm behind}$}}

\begin{figure}[!ht]
\center{\includegraphics[width=1\textwidth]{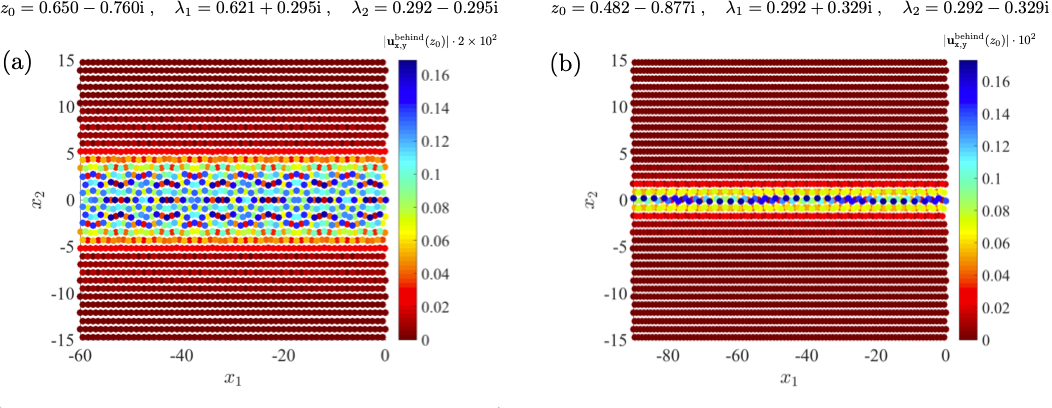}}

\caption{Modes produced by a heavy defect ($\beta=1$), interacting with an incoming wave with the frequency $\omega=1$, amplitude $0.01$ and angle of incidence $47^\circ$ measured anticlockwise from the positive $x_1$-axis. We show localised defect modes corresponding to (a) a longitudinal wave and (b) a transverse wave propagating along the defect. These illustrations have been performed based on (\ref{ubehind}).}
\label{pic_mode_ahead_2}
\end{figure}

In Figure \ref{pic_mode_ahead_2}, we demonstrate the modes ${\bf u}_{{\tt x},{\tt y}}^{\rm behind}$ in (\ref{ubehind}) for the same set-up as associated with Figure \ref{pic_mode_ahead}. Figure \ref{pic_mode_ahead_2}(a) shows an example of the localised defect mode corresponding to the longitudinal vibration of the defect. There, the defect is subjected to a tensile-type dynamic deformation that induces the symmetric response encountered in the class $\mathcal{T}$. This response is localised with respect to the direction perpendicular to the defected lattice row, which is in correspondence with the values of $\lambda_{1,2}$ shown in the Figure. Likewise, a transverse vibration mode induced by the dynamic shearing for the same defect and loading configuration is demonstrated in Figure \ref{pic_mode_ahead}(b), where we observe a clear anti-symmetric pattern in the response of the system behind the defect tip. The modes shown in Figure \ref{pic_mode_ahead_2} combine with those in Figure \ref{pic_mode_ahead} and other modes in the medium to retrieve the far field lattice response of the system (see Figure \ref{pic_beta_1}(a) in Section \ref{sec5}). The knowledge of the combination of such modes and the characterisation of the behaviour of the field near the defect tip is fully attributed to the Wiener-Hopf technique and its application developed here.

Below, we analyse the dispersive properties of the system that lead to wave modes ${\mathbf{u}}^{\text{behind}}_ {{\tt x}, {\tt y}}$ generated in the bulk when the inertial defect vibrates.

\subsubsection{Dispersive properties of the inertial defect}\label{sec3p5p2}
We consider three spectral regimes defining frequencies where the waves (\ref{ubehind}), corresponding to $|z_0|=1$, can propagate along the inertial defect. These regimes include:
\begin{enumerate}[$\bullet$]
\item {\bf Regime I}, described by the frequency range $\omega \ge 2.25$, that coincides with the support of both high-frequency transverse and longitudinal waves along the inertial defect.
\item {\bf Regime II}, which is contained in an intermediate frequency interval approximately given by ${1.732}\le \omega \le {2.236}$. In this case, the structured inertial defect can support either longitudinal or transverse motion for heavy or light inertial defects, respectively.
\item {\bf Regime III}, situated in the low-frequency range $\omega \le \sqrt{2}$, where both longitudinal and transverse wave motion along the heavy inertial defect can be supported or these inertial defect wave modes can be excited separately. 
\end{enumerate}
\begin{figure}[!ht]
\center{\includegraphics[width=\textwidth]{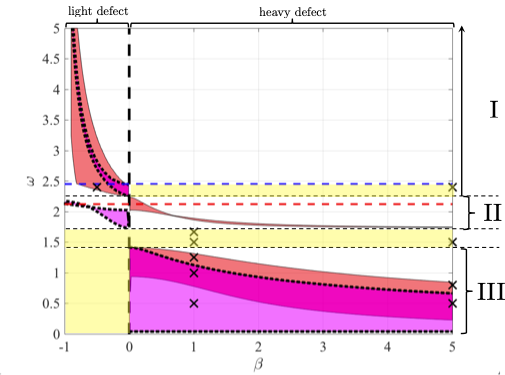}}
\caption{Pass band structure of the inertial defect as a function of $\beta$ based on (\ref{poly1}) and (\ref{poly2}) in Appendix \ref{zeros}. The vertical width of the magenta (red) regions bounded by black curves (dashed curves) indicate the pass band width for transverse (longitudinal) wave motion of the inertial defect characterised by $\beta$. {Those bounding curves (shown after interpolation) have been identified numerically by finding the minimum and maximum frequencies for which solutions with $|z|=1$ of (\ref{poly1}) and (\ref{poly2}) exist for each value of $\beta\ne0$.} The blue horizontal dashed line represents the frequency upper bound $\omega=\sqrt{6}$ for the pass band of the incident waves (see Figure \ref{pic_dd_bulk}). The red horizontal dashed line is the cut-off frequency for $\omega_-$ in (\ref{incdisp}). Here, the black horizontal dashed lines are the bounding frequencies for Regimes I--III. The vertical dashed line is connected with special case $\beta=0$ when the lattice is uniform. Yellow rectangles are parameter subdomains where the incident wave does not excite wave motion along the inertial defect for any $\omega$ and $\beta$. The black crosses correspond to the values of $\omega$ and $\beta$ for the computations of the scattered field presented in this article given by (\ref{UxyR}) (see Section \ref{sec5}). }
\label{pic_dd_band}
\end{figure}
These three frequency regimes are indicated on Figure \ref{pic_dd_band}, which shows frequencies for when longitudinal and transverse inertial defect waves can exist as a function of the defect's inertial properties $\beta$. This has been determined by analysing the behaviour of the zeros for $\mathfrak{L}_1(z)$ and $\mathfrak{L}_2(z)$ representing these tensile and shear wave modes produced by the inertial defect, respectively, that reside along the unit disk in $\mathbb{C}$. This analysis reduces to the solution of high-order polynomials in $z$, which are presented in the Appendix \ref{zeros} (see (\ref{poly1}) and (\ref{poly2})).

A global conclusion concerning the dynamic response of the defect, taken from Figure \ref{pic_dd_band} and expected on physical grounds, is that the light (heavy) defects require waves of sufficiently high (low) frequencies to be excited.
Interestingly, in transitioning from the low to high frequency regimes, Figure \ref{pic_dd_band} also demonstrates that Regime II acts as the intermediate zone where the defect has a preference to support either a transverse mode in the case of heavy defect or a longitudinal mode for the light defect.

Finally, while the current study concerns scattering of elastic waves by a structured inertial defect, we note the information presented in Figure \ref{pic_dd_band} can be used to study highly localised lattice waves generated along the inertial defect. In this context, Figure \ref{pic_dd_band}, shows it is possible to generate these wave modes for frequencies in the range $\omega>\sqrt{6}$, provided the inertial defect is light, i.e. $-1<\beta<0$
and this is shown in Figure \ref{pic_dd_bulk} as the horizontal red-dashed line 
shown as the horizontal blue line in Figure \ref{pic_dd_band}.

\subsubsection{Pass bands for the semi-infinite inertial defect}
\label{lightheavydefect}
We investigate the behaviour of the pass bands for the structured inertial defect characterised by $\beta$. We carry this out in accordance with the approach taken in Section \ref{sec_disp_bulk} and in connection with Figure \ref{pic_dd_band}.

\subsubsection*{The light inertial defect}
Figure \ref{pic_dd_band} shows when $-1<\beta <0$, the light inertial defect possesses pass bands in Regimes I and II. Note that the light inertial defect does not support wave modes outside these regimes. 

\begin{figure}[!ht]
\center{\includegraphics[width=0.8\textwidth]{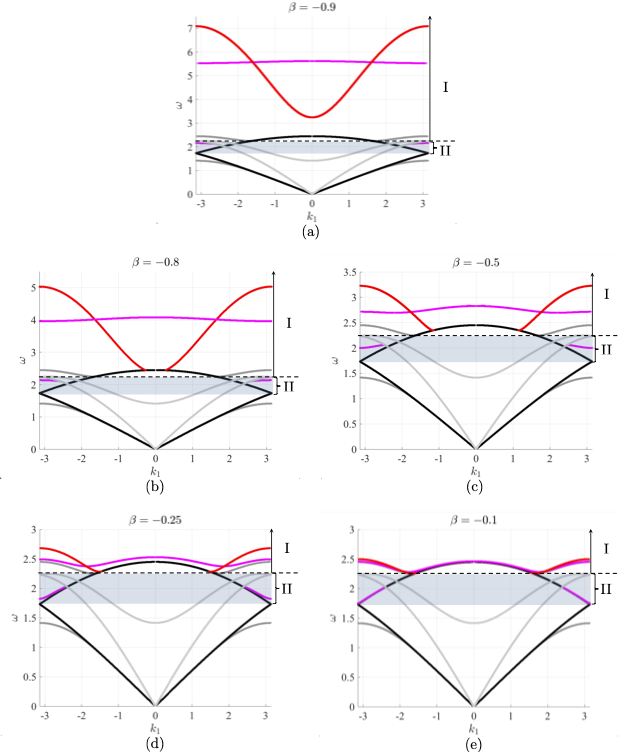}}
\caption{Dispersion curves for the light inertial defect defined by $-1<\beta<0$. Regime I, bounded below by the horizontal dashed line, and Regime II (indicated by the blue shaded frequency interval) are defined to the right of each diagram. Red (magenta) dispersion curves are those corresponding to longitudinal (transverse) wave motion supported along the inertial defect based on solutions to (\ref{poly1}) and (\ref{poly2}), respectfully, with $|z|=1$. The remaining curves are those for the bulk medium (see Figure \ref{pic_dd_bulk} for their description). }
\label{pic_dd_b_l_0}
\end{figure}

In Regime II, only pass bands for the transverse wave modes of the inertial defect are present. These bands are defined for discrete intervals of the wavenumber and these intervals are extremely narrow for approximately $\beta \le -0.65$, indicating the associated dispersion curves are almost flat (see the magenta curves in Figures \ref{pic_dd_b_l_0}(a) and (b), for example). For light inertial defects with more inertia, Figure \ref{pic_dd_band} shows the pass band width for transverse wave modes in Regime II can be larger and this band is located at lower frequencies. This is evidenced in Figures \ref{pic_dd_b_l_0}(c) and (d), where it is clear that as $\beta \to 0$, the magenta dispersion curve tends to the lowest curve associated with symmetry $\mathcal{S}$ bulk lattice motion in Regime II (see Figure \ref{pic_dd_b_l_0}(e)). This behaviour in the topology of the band in reaching the limit $\beta \to -0$ is consistent physically with the making the inertial defect heavier until a lattice with uniformly distributed inertia is reached.

Figure \ref{pic_dd_band} also demonstrates that in the high frequency Regime I for $\beta<0$, transverse inertial defect motions are generally supported in narrow frequency intervals. In addition, these mechanical oscillations are always accompanied by the longitudinal wave modes. These waveforms are supported in wider frequency intervals compared with those where transverse waves can be excited. An example demonstrating these observations is given in Figure \ref{pic_dd_b_l_0}(a) and (b), where the magenta dispersion curves for transverse wave motion are almost flat in Regime I, whereas the red curves for longitudinal wave motion along the inertial defect occupy a broad range of high frequencies. 
 In connection with this, as $\beta \to -1$, Figure \ref{pic_dd_band} indicates the pass band for transverse (longitudinal) wave motion along the inertial defect shrinks (expands) in Regime I.

 When $\beta \to 0^-$, Figure \ref{pic_dd_band} and Figures \ref{pic_dd_b_l_0}(a)--(e) show the pass bands for longitudinal and transverse inertial defect wave motion in Regime I coalesce. As this happens, the inertial defect's dispersion curves also merge on to the upper most dispersion curves for the bulk lattice (see Figure \ref{pic_dd_b_l_0}(e)). Further, as $\beta \to 0^-$, the band for longitudinal wave motion in Regime I and the pass band for the bulk medium transition from being disjoint to interacting with each other. This starts from approximately $\beta=-0.82$ where the bulk dispersion curves and those for the longitudinal motion of the inertial defect coincide at the point $(k_1, \omega)=(0,\sqrt{6})$. The interaction of these bands for $\beta> -0.82$ leads to the highest bulk lattice dispersion curves controlling the lowest frequency for the tensile lattice wave modes produced by the inertial defect in Regime I. Additionally, we note that this interaction promotes the appearance of longitudinal inertial defect modes defined for disjoint intervals of the wavenumber that shrink as $\beta \to 0^-$. This is demonstrated in Figure \ref{pic_dd_b_l_0}(b)--(e), where the dispersion curves for the longitudinal inertial defect modes are defined in a piecewise manner with respect to $k_1$ (see also \cite{Piccolroazetal} where a similar phenomenon can be observed in the Mode III fracture of a dissimilar lattice medium). Alternatively, for $-1<\beta<0$, the dispersion curve for transverse wave motion along the inertial defect remains continuous as shown in examples of Figures \ref{pic_dd_b_l_0}(a)--(e).
 
 Finally, in the context of wave scattering, when $0>\beta >-0.82$ we note there exist frequencies in Regime I where the incident wave can excite wave motion along the inertial defect. According to Figure \ref{pic_dd_band}, this occurs for approximately $2.25 < \omega < 2.45 \approx \sqrt{6}$, where inertial defect can support longitudinal waves with (for $-0.82 >\beta>-0.33$) or without (for $-0.33<\beta<0$) the presence of transverse waves.
 
\subsubsection*{The heavy inertial defect}
\begin{figure}[!ht]
\center{\includegraphics[width=1\textwidth]{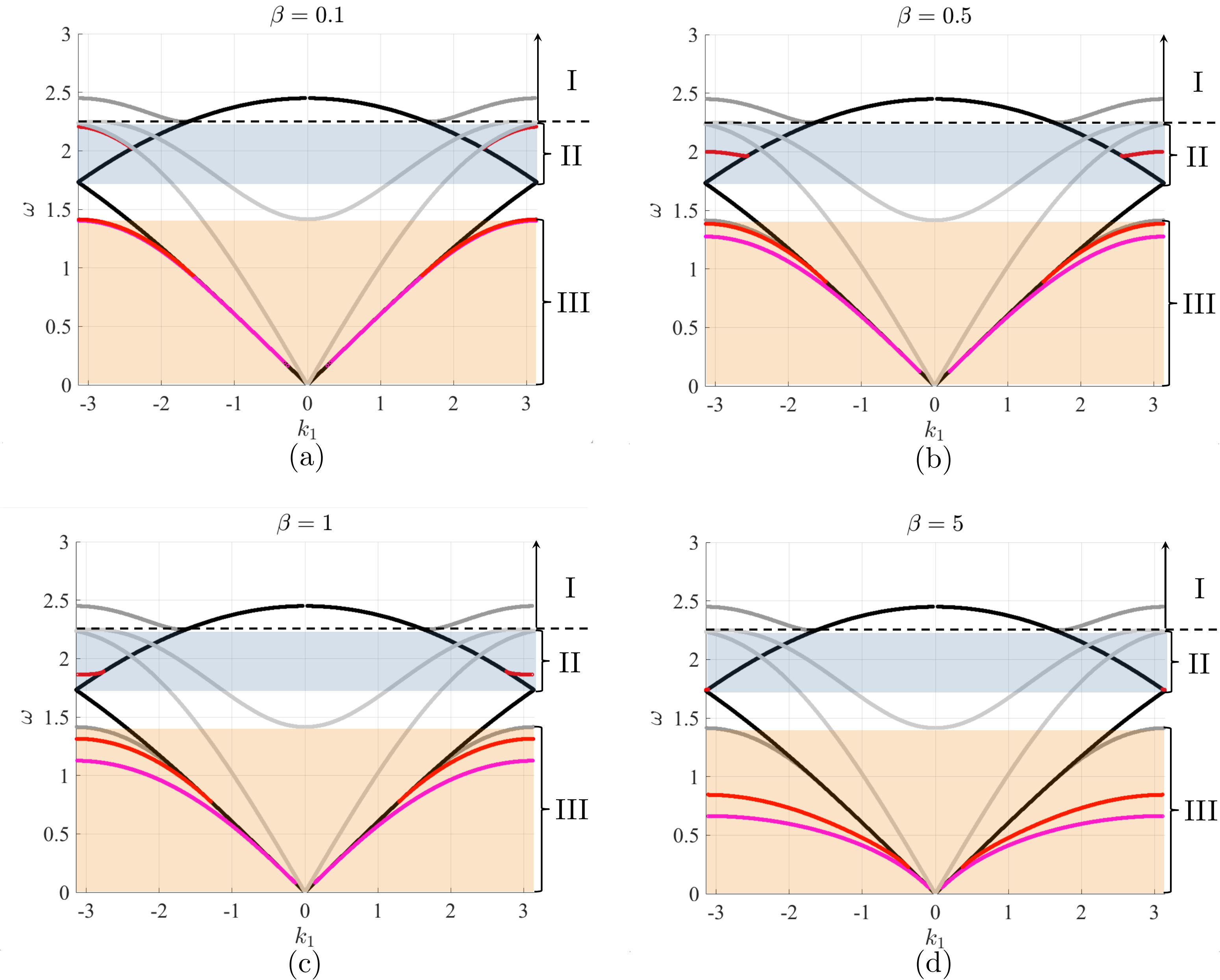}}
\caption{Dispersion curves for the heavy inertial defect described by $\beta>0$. Region III is shown as the orange shaded frequency interval. We refer to {Figure \ref{pic_dd_bulk} and} Figure \ref{pic_dd_b_l_0} for further descriptions. }
\label{pic_dd_1_b_g_0}
\end{figure}

When $\beta>0$, only wave motion of the heavy inertial defect is excited for frequencies in Regimes II and III, as shown in Figure \ref{pic_dd_band}. No waves are supported by the heavy inertial defect for frequencies outside these regimes. At a narrow range of frequencies in Regime II, the inertial defect only supports longitudinal wave motion in connection with real wavenumbers belonging to a discrete periodic collection of sets (see Figures \ref{pic_dd_1_b_g_0}(a)-(d)). In particular, Figure \ref{pic_dd_band} shows the inertial defect's capacity to sustain these wave modes in Regime II is severely reduced for approximately $\beta >0.67$, where the associated pass band becomes very narrow. Correspondingly, the dispersion curve defined on these intervals is almost flat (see examples in Figures \ref{pic_dd_1_b_g_0}(b) and (c)). 

Further, similar to the phenomena observed for the light inertial defect in Regime I, there is interaction between the dispersion curves for the bulk and longitudinal inertial defect modes in Regime II. This occurs for $\beta>0.67$ and the phenomenon is associated with increase of the inertial properties of the inertial defect. In Regime II, the black dispersion curve for symmetry $\mathcal{S}$ wave modes in the bulk interacts with the red dispersion curve for the inertial defect as it moves down in the frequency spectrum for increasing $\beta$. This results in a reduction in the width of the wavenumber intervals for which the corresponding longitudinal inertial defect mode is defined. Further, as the inertia of the defect increases, these intervals tend to points at the lowest frequency inside Regime II, as demonstrated for $\beta=5$ in Figure \ref{pic_dd_1_b_g_0}(d). 

In contrast to the frequency Regime I for light inertial defects, within certain frequency intervals in Regime III, the inertial defect can support both longitudinal or transverse waveforms or may support only one type of these waves. Additionally, transverse inertial defect wavemodes can be excited at a wider range of frequencies in comparison with those associated with longitudinal waves, which differs from the effect observed in Regime I for the light inertial defect where opposite scenario occurs).

\section{Numerical illustrations}\label{sec5}

We use the analytical solution (\ref{UxyR}) to compute the scattered field for the considered problem and we validate our approach with a comparison against a benchmark numerical approach in COMSOL Multi-Physics.

\subsection{Computational settings and the finite element model}
The problem for the scattered field is programmed in the Structural Mechanics module of COMSOL Multi-Physics 5.3. For this, we utilise a slab of triangular lattice with the dimensions $120\times 121 \sqrt{3}/2$, corresponding to the lattice with 120 lattice spacings (taken as unity) horizontally and 120 rows. A finite length defect is included in the computational geometry along $n_2=0$ (corresponding to the 61st row), within the interval $-60 \le x_1<0$ at $n_2=0$ (see Section \ref{sec2} for the definition of $x_j$, $j=1,2$). The load produced by the incident field in the considered theoretical problem is programmed along the nodes forming the defect. The incident wave is normalised such that $|{\bf a}^{\mathfrak{i}}|=0.01$, i.e. $a$ in (\ref{ainc}) selected appropriately to enable this. The frequency $\omega$ is selected within the interval $0<\omega<\sqrt{6}$ and the normalised wave numbers ${k}_{\tt x}$ and ${k}_{\tt y}$ are determined via 
 (also recall \eqref{k1k2}, \eqref{k1k2vec})
\begin{equation}\label{modktrig}
 {k}_{\tt x}=2^{-1}|{\bf k}|\cos\Theta \quad \text{ and } \quad {k}_{\tt y}=2^{-1}\sqrt{3}|{\bf k}|\sin\Theta\;,
 \end{equation}
with $\Theta$ being the prescribed angle of incidence and $|{\bf k}|$, with ${\bf k}=(k_1,k_2)^{\rm T}$, being found from the degeneracies of (\ref{ID}) corresponding to its non-trivial solutions.
The lattice slab is also supplied with an Adaptive Absorbing Layer (AAL) in the immediate location of the exterior boundary and having a width of 10 lattice spacings to help in minimising the reflections produced by the exterior of the slab.

\begin{figure}[!ht]
\center{\includegraphics[width=0.7\textwidth]{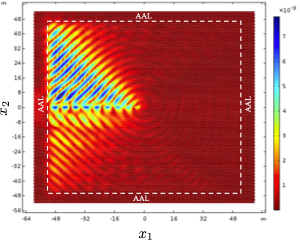}}
\caption{Finite element computation for the scattered field in COMSOL Multi-Physics 5.3. The interior boundary of the AAL is indicated by the white dashed line. The color bar indicates the total displacement within the structure.}
\label{pic_fe_1}
\end{figure}

Figure \ref{pic_fe_1} provides the total displacements for the scattered field in the lattice resulting from the interaction of the defect and an incident wave with a low frequency, $\omega=0.5$, and angle of incidence with $\Theta=227\pi/180 \text{ rad}$, i.e. $47^\circ$ measured anti-clockwise relative to the positive $x_1$-axis. The wavenumber is defined using $\omega_-$ in (\ref{incdisp}), which at low frequencies defines shear waves analogous to those observed in the analogous continuum model.

\begin{figure}[!ht]
\center{\includegraphics[width=1\textwidth]{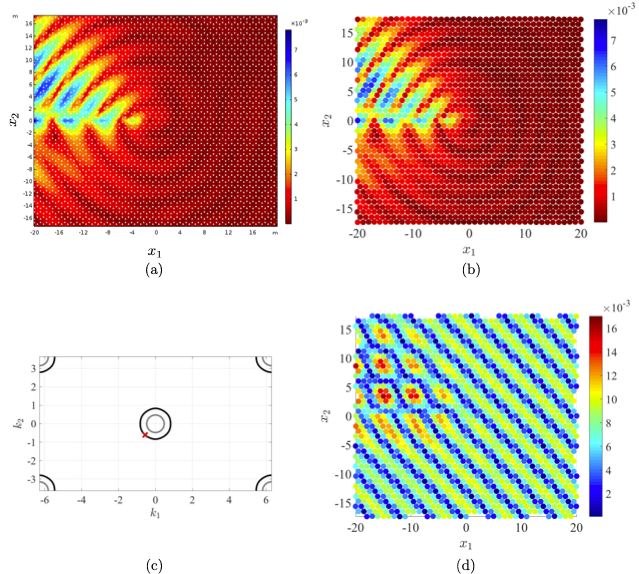}}
\caption{Comparison of computations for the scattered field induced by a shear wave in the vicinity of the defect edge {in (a) and (b)}. The COMSOL Multi-Physics computation is shown in (a) and whereas the scattered field computed via (\ref{UxyR}) is given in (b). Slowness contours for the bulk lattice at $\omega=0.5$ based on (\ref{incdisp}) {are provided in (c)}. The red cross shown indicates the wavenumbers $k_1$ and $k_2$ defining the incident wave (see (\ref{dim5}) and (\ref{modktrig})). The total field based on the analytical solution (\ref{incscasum}), (\ref{uinc}) and (\ref{UxyR}) is provided in {(d)}. {The color bars in this Figure indicate the total displacement of the medium.} }
\label{pic_comparison_1}
\end{figure}

The defect is heavy and is formed from nodes having twice the mass of the ambient lattice nodes (i.e. $\beta=1$). This example corresponds to a case where the defect can support a localised transverse wave mode (see Figure \ref{pic_beta_1}). Accordingly, there is an oscillation observed along, approximately, $-50<x_1<0$, $x_2=0$ in this finite element computation.
As a demonstration of the accuracy of the developed approach, Figure \ref{pic_comparison_1}(a) and (b) show the field in the vicinity of the defect edge computed in COMSOL and via (\ref{UxyR}), respectively. There is a very good agreement between both predictions for the scattered field. Both computations illustrate that the defect tip behaves as a source producing outgoing waves with circular fronts. This is further supported by the slowness contours in Figure \ref{pic_comparison_1}(c) for the bulk medium that are circular for the low-frequency considered, akin to those found in the continuum approximation discussed in 
Section \ref{inc_w_sec}. Additionally, the defect reflects some of the incoming wave energy quadrant of the lattice and these reflected waves propagate above and behind the defect edge.
 Some of the incident wave energy also penetrates the defect. The latter effect is illustrated in Figure \ref{pic_comparison_1}(d), where perturbation to the incident wave field can be observed below the line defect in the third lattice quadrant.

We also observe discrepancies between the finite element computations and those based on the analytical solution in Figures \ref{pic_comparison_1}(a) and (b). They can be attributed to the AAL applied in the finite element model, which is sensitive to the frequency and geometry of the lattice and this can be time consuming to optimise. Indeed, despite the tuning of the AAL to minimise reflections from the exterior boundary, Figure \ref{pic_fe_1} shows small aberrations in the field can occur that are visible in regions where the scattered field magnitude is small. These can also impact the near field of the defect edge. In comparison, the analytical solution based on the Wiener-Hopf technique is easily implemented to accurately recover the features of the scattered field.

\subsection{Scattering phenomena for the linear elastic defect}

\begin{figure}[!ht]
\center{\includegraphics[width=1\textwidth]{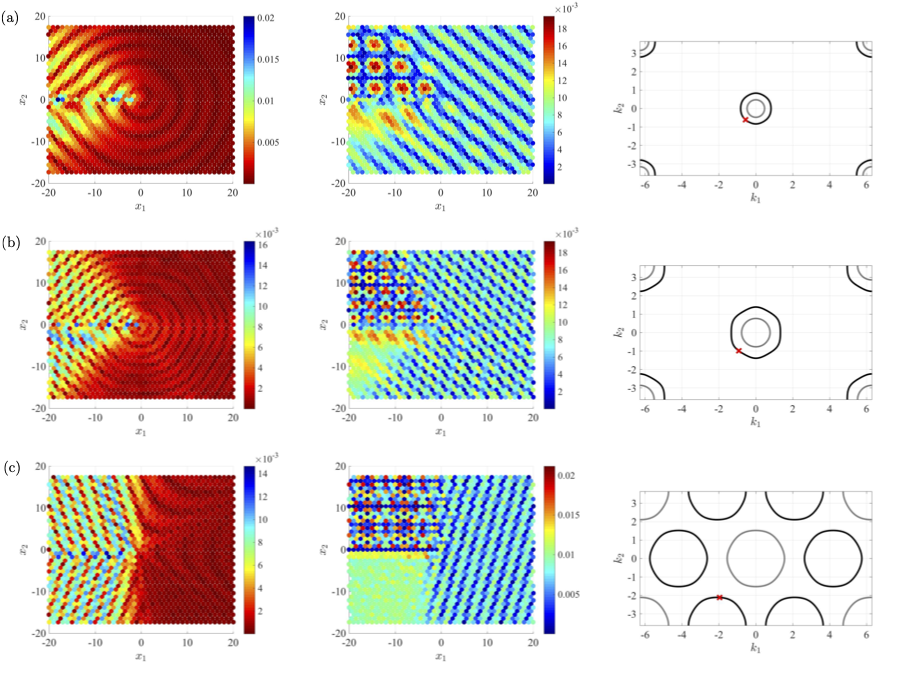}}
\caption{The response of the lattice having a heavy defect, with $\beta=5$, and supporting an incident wave defined with $\omega_-$, $\Theta= 227 \pi/180 \,{\rm rad}\, (\equiv 47^\circ$ measured anticlockwise from the positive $x_1$-axis) and frequency (a) $\omega={0.5}$, (b) {$\omega=0.8$} and (c) $\omega=1.5$. {The computations for the total displacement, indicated by the color bars,} have been performed with (\ref{incscasum}), (\ref{uinc}) and (\ref{UxyR}). Left (central) panels show the total displacements for the scattered (total) field. The right panels are the slowness contours for the specified frequency using (\ref{incdisp}), with grey (black) contours being those given by $\omega_+$ ($\omega_-$) in (\ref{incdisp}). The red cross represents the wavevector ${\bf k}$ of the incident wave for each frequency.}

\label{pic_beta_5}
\end{figure}

Figures \ref{pic_beta_5} and \ref{pic_beta_1} show the response of the lattice subjected to an incident wave, defined using $\omega_-$ in (\ref{incdisp}) and $\Theta=227\pi/180 \text{ rad}$, for varying frequencies and different heavy defects. On the other hand, Figure \ref{pic_beta_m0p5} includes the scattering induced by both light and heavy defects in the case of a high frequency incident wave based on $\omega_+$, with $\Theta=7\pi/6 \text{ rad}$.

{
In all cases presented, the defect edge acts as a source radiating waves into the bulk lattice owing to its interaction with incident wave (see the left panels in Figures \ref{pic_beta_5} and \ref{pic_beta_1}). Figures \ref{pic_beta_5}(a) $\&$ (b) and \ref{pic_beta_1}(a) $\&$ (b) all correspond to cases where the incident wave is able to excite wave propagation along the defect (see Figure \ref{pic_dd_band}). At low frequencies, the dynamic behaviour of the bulk lattice is isotropic. Indeed, as mentioned above, in this frequency range the lattice approximates an elastic continuum (see Section \ref{inc_w_sec} and \cite{slepyansbook}), where waves are non-dispersive and scattered waveforms radiating from the defect tip have circular fronts (e.g. as in Figure \ref{pic_comparison_1}(a) \& (b) and the left panel in Figure \ref{pic_beta_5}(a)). 
Further, owing to the dynamic anisotropy of the structured medium, Figure \ref{pic_beta_5}(b) indicates the wave field radiating from the edge can possess an almost hexagonal wavefront. Moreover, increasing the incident wave frequency leads to clear preferential directions emerging in the scattered bulk response. This effect can produce highly localised waves radiating from the defect edge, at angles $60^{\circ}$ and $120^{\circ}$ measured anti-clockwise from the positive $x_1$-axis, as shown in Figure \ref{pic_beta_1}(a) and (b). It also enables scattered star-shaped wavefronts \cite{SlepyanAyzenbergStepanenko} as revealed in Figures \ref{pic_beta_5}(c), \ref{pic_beta_1}(c). We note that the observed effects are consistent with the slowness contours shown in the right panel of Figures \ref{pic_beta_5} and \ref{pic_beta_1}.
In Figure \ref{pic_beta_1}(c), there is also clear evidence in the first lattice quadrant that the defect enables mode conversion during the wave scattering process. This allows incident waves with short wavelengths to be scattered as waveforms with larger wavelengths. At low frequencies this corresponds to the conversion of shear waves to pressure waves in the lattice. We report that it is also possible to find the opposite effect in the considered system.

In the low-frequency regime, in correspondence with scattering effects produced by semi-infinite screens in continuum models, Figures \ref{pic_beta_5}(a) \& (b) and Figure \ref{pic_beta_1} show that a portion of incident wave is reflected (transmitted) by the defect into the second (third) lattice quadrant.
On the other hand, Figure \ref{pic_beta_1}(d) demonstrates the dynamic anisotropy of the lattice at higher frequencies enables the incident wave to undergo negative reflection, where reflected waveforms enter the first quadrant of the lattice. Simultaneously, Figure \ref{pic_beta_1}(d) illustrates the incoming wave experiences negative refraction upon penetrating the defect leading to the propagation of transmitted waves into the fourth lattice quadrant.

\begin{figure}[htbp]
\center{\includegraphics[width=1\textwidth]{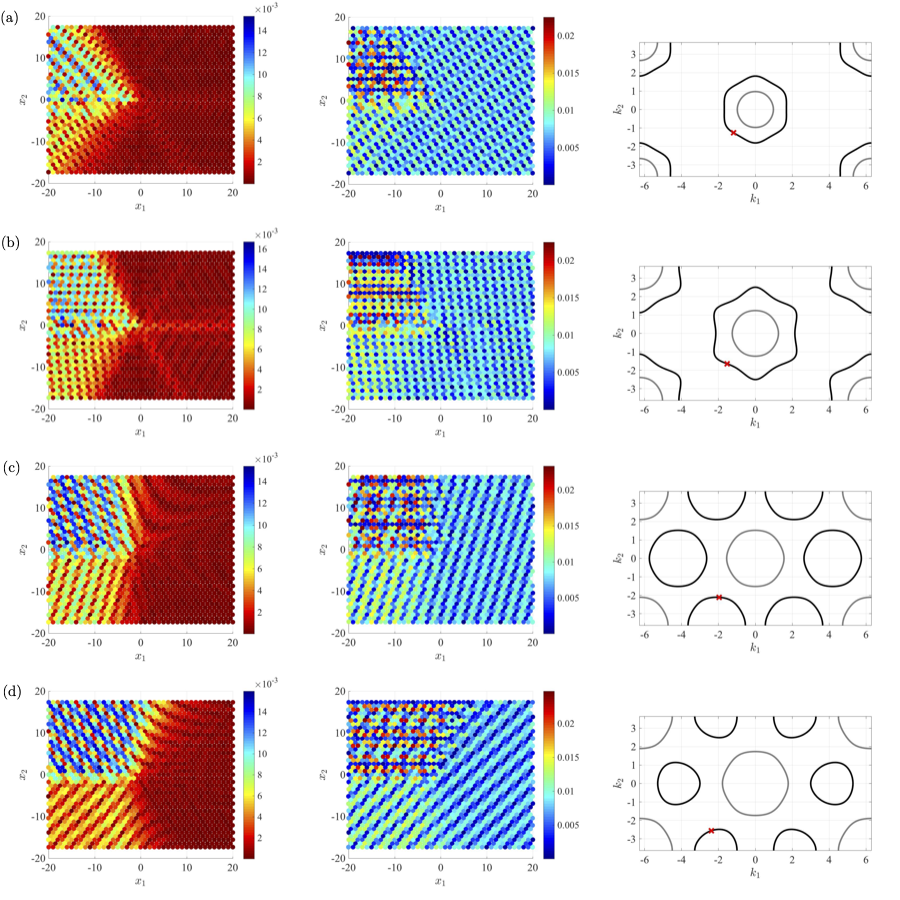}}
\caption{The response of the lattice containing a heavy defect with $\beta=1$ and supporting an incident shear wave with $\Theta= 227 \pi/180 \text{ rad}$ and frequency (a) $\omega={1}$, (b) {$\omega=1.25$}, (c) $\omega=1.5$ and (d) $\omega=1.67$. The computations are based on (\ref{incscasum}), (\ref{uinc}) and (\ref{UxyR}). See Figure \ref{pic_beta_5} for the description of the left, central and right panels in (a)-(c).}
\label{pic_beta_1}
\end{figure}

Since the defect is movable, it also plays a role in transmitting energy that can impact the response of the ambient lattice, as demonstrated in Figures \ref{pic_beta_5} and \ref{pic_beta_1}. They show the strength of this transmitted field, found below the defect, is clearly sensitive to the incident wave frequency and inertial properties of the defect. {When the defect can vibrate, admitting a localised bulk mode, this can provide some enhancement the magnitude of the scattered field in the vicinity of the defect. }

\begin{figure}[!ht]
\center{\includegraphics[width=1\textwidth]{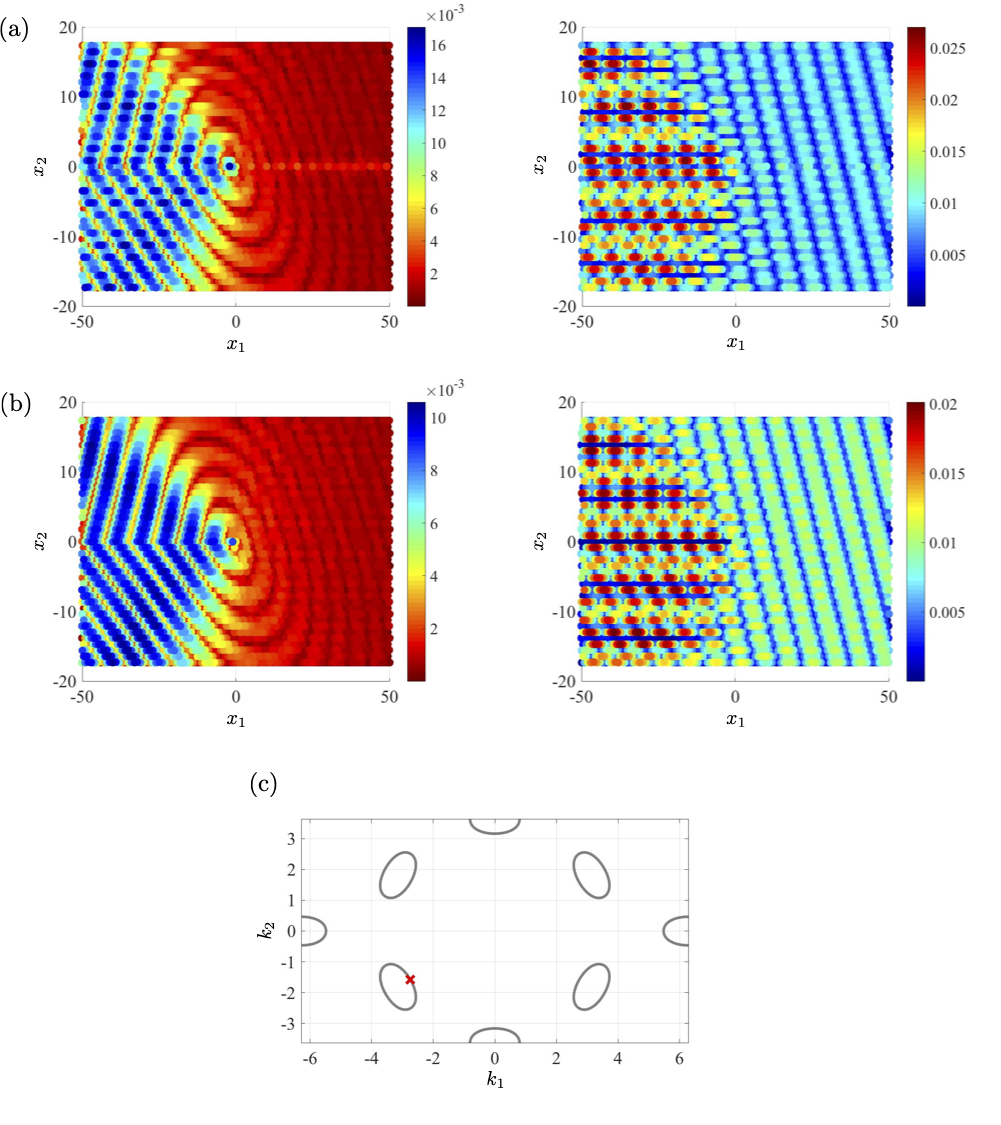}}
\caption{
{The response of the lattice containing different structured defects that interact with an incoming wave with frequency ${\omega} =2.4 $ and angle of incidence at $\Theta=\pi/6$ rad measured anti-clockwise from the positive ${x}_1$-axis. 
In (a) the defect is light with $\beta=-0.5$, and in (b) $\beta=5$, where the defect is considered as heavy. In (a) and (b) the left (right) panels show the total displacement (indicated by the color bar) of the lattice nodes corresponding to the scattered (total) fields. Tehey have been computed with (\ref{incscasum}), (\ref{uinc}) and (\ref{UxyR}). In (c) the slowness contours (grey curves) for the medium ambient to the defect is shown. There, the red-cross corresponds to the wave vector defining the incident wave.
}}
\label{pic_beta_m0p5}
\end{figure}

As discussed in Section \ref{sec3p5p2}, and as expected on physical grounds, heavy (light) defects are likely to be excited by incident waves with low (high) frequencies and long (short) wave lengths. In relation to this, Figure \ref{pic_beta_m0p5} demonstrates the scattering of high-frequency waves by both light and heavy defects in the case when the incoming wave is defined using the dispersion surface $\omega_+$. In particular, at the chosen frequency the scattering process only initiates symmetry $\mathcal{S}$ and mixed modes (see Section \ref{symmsec}). Further, the dynamic behaviour of the bulk lattice is highly anisotropic, as indicated the slowness contours in Figure \ref{pic_beta_m0p5}(c) represented by small ellipses centred at on the vertices of a regular hexagon of side length $\pi$. In the computations of the scattered field in Figure \ref{pic_beta_m0p5}(a) and (b), the defect radiates waves whose wave fronts are elliptical. We note that there is a clear disparity between the maximum total displacement in each of these computations, with Figure \ref{pic_beta_m0p5}(a) coinciding with the case when a longitudinal wave is initiated along the defect by the incident field. In correspondence with this effect, we observe defect vibrations clearly induce the propagation of wave modes ahead of the defect along $x_2=0$. This is evidenced by the highly localised oscillations in the scattered and total fields in Figure \ref{pic_beta_m0p5}(a) along the line containing the defect. Further, the scattered field in Figure \ref{pic_beta_m0p5}(a) shows the localised dynamic modes of the defect create oscillations in the reflected and transmitted components of the incident field located in the second and third lattice quadrants, respectively. In contrast, in the case when the defect is heavy, the same incident wave does not excite the defect (see Figure \ref{pic_beta_m0p5}(b)). Consequently, those localisation effects and bulk waves produced by the defect in Figure \ref{pic_beta_m0p5}(a) are no longer observed. Further,
it apparent that the total field in Figure \ref{pic_beta_m0p5}(b) shows the defect appears to be undisturbed by the incident wave.

\section{Conclusions}\label{sec6}
The scattering of in-plane elastic waves by a linear inertial defect in an elastic triangular lattice has been studied. While the governing equations for this lattice are naturally described as a system where principal displacements of each lattice node are coupled, the analysis of the corresponding discrete Fourier transformed equations and inherent symmetries of the problem have enabled us to simplify the problem dramatically, providing a scalar Wiener-Hopf equation describing the lattice behaviour.

Consequently, the standard method of solution \cite{Noble, Bls1} to such equations has provided a compact and exact representation of the lattice displacements in the form of a contour integral.
Asymptotic analysis of the corresponding integral kernel has revealed that in the remote lattice, i.e. far from the defect edge, the lattice response involves at most three classes of dynamic motions. These classes involve fundamental modes of the system and the existence of two of these sets is frequency dependent. One such class includes a set of axi-symmetric modes, corresponding to a dynamic shearing acting parallel to the defect. Those modes are present in the scattering of any incident wave. The other classes accompanying these remote shear-type modes include those supported inside subsets of the bulk lattice pass band and identified as (i) dynamic tensile modes, supported within zero-frequency pass band, 
and (ii) a class of modes involving a mixture of dynamic shear and tensile motions that occur in two disjoint high-frequency pass bands.

Further, inspection of the kernel function corresponding to the Wiener-Hopf equation has enabled the complete dynamic characterisation of the scattering response of the medium for any defect and incident wave frequency. This follows from analytical dispersion relations, appearing as polynomials involving transform parameter. They provide for the predictions of scattering regimes where interfacial evanescent and propagating modes are supported along the defect, in addition to those oscillating and evanescent modes that decay in far-field of the bulk lattice.

The numerical computations that we have presented demonstrate that the analytical method leads to an accurate solution that compares well with an independent solution calculated via finite element method. Contrary to the analytical formulation of the problem considered, the finite element solution requires a bounded computational domain and tuning of damping layers that minimise reflections from the exterior of this domain.
On the other hand, our approach can be directly implemented in a numerical schemes and allows for the direct computation of the scattered field at any given lattice node.

Further, the numerical investigation of the scattering effects produced by the structured defect has enabled the full characterisation of the response of the system for a broad range of frequencies. It also provides insight into how the lattice imitates the classical behaviour observed in analogous continuum scattering problems at low frequencies. This is in addition to demonstrating how special effects such as dynamic anisotropy, negative reflection and refraction attributed to the periodicity of the underlying structure can be engaged for systems whose geometry abruptly change.

The work presented in this article contributes to research into analytical approaches for modelling elastic wave motion for vectorial problems involving periodic engineering systems. It is envisaged that the method presented will naturally lead to solution of analogous problems posed for more exotic structures capable of special responses. It is anticipated work presented will also have applications in civil engineering where the analysis of wave motion in complex structures in society is of paramount importance, in addition to the design of carbon nanotubes whose exceptional strength make them extremely useful in developing novel composite materials and nanotechnologies.

\section*{Acknowledgement}
{The authors thank the anonymous reviewers for their constructive comments and suggestions.}
M.J.N. gratefully acknowledges the support of the EU H2020 grant MSCA-RISE-2020-101008140-EffectFact. The authors would also like to thank the Isaac Newton Institute for Mathematical Sciences (INI) for their support and hospitality during the programme – ‘Mathematical theory and applications of multiple wave scattering’ (MWS), where work on this paper was undertaken and supported by EPSRC grant no EP/R014604/1. Additionally, the authors are grateful for the funding received from the Simons Foundation that supported their visit to INI during January-June 2023 and participation in MWS programme.
Last but not the least, the authors are also grateful to the fruitful discussions on the scattering theory with the participants of workshops at the MWS programme which has influenced the literature survey.

\bibliographystyle{plainnat}

\appendix

\section{Transformed solution for the bulk lattice}\label{Uupper}
We provide some details of the derivation of the representation (\ref{u_upper}) of the solution to (\ref{bulk_eq_tran}) for ${\tt y}\in{{\mathbb{Z}}_+}\setminus\{0\}$. We look for the vector function ${\mathbf{u}}^{\rm F}_{{\tt y}}$ in the form
\begin{equation}\label{usol_outside}
{\mathbf{u}}^{\rm F}_{{\tt y}}={\bf p}(\alpha) \alpha^{\tt y}\;, \quad \text{ with } |\alpha| < 1\;, \quad \text{ for } {{\tt y}} \in{{\mathbb{Z}}_+}\setminus\{0\}\;.
\end{equation}
Here the condition on $\alpha$ ensures a physically reasonable solution for the scattered field that decays sufficiently fast in the remote lattice.
In this case, the transformed equations (\ref{bulk_eq_tran}) for the total field in the bulk lattice are written as
\begin{eqnarray}
- \omega^2{\bf p}(\alpha)&=& {\bf a}^{(1)}\cdot {\bf p}(\alpha)\left[z^2+z^{-2}-2\right]{\bf a}^{(1)}\nonumber \\
&&+{\bf a}^{(2)}\cdot {\bf p}(\alpha) \left[\alpha z^{-1}+\alpha^{-1}z-2\right]{\bf a}^{(2)}\nonumber\\
&&+{\bf a}^{(3)}\cdot {\bf p}(\alpha) \left[\alpha z+\alpha^{-1} z^{-1}-2\right]{\bf a}^{(3)}\;.
\end{eqnarray}
With this, we arrive at the homogeneous system:
\begin{equation}\label{eqc}
({\bf C\rm}({z,}\alpha)+ \omega^2{\bf I\rm}){\bf p}(\alpha)={\bf 0\rm}
\end{equation}
where 
\begin{equation}
{\bf C \rm}({z,}\alpha)=\left( \begin{array}{cc}
\frac{1}{4}\, \left( \alpha+{\alpha}^{-1} \right) \left( z+{z}^{-1}
 \right) + \left( z+{z}^{-1} \right) ^{2}-5
 & \frac{\sqrt{3}}{4} (\alpha-{\alpha}^{-1})(z-{z}^{-1})\\ \\
\frac{\sqrt{3}}{4} (\alpha-{\alpha}^{-1})(z-{z}^{-1}) & \frac{3}{4}\, \left( \alpha+{\alpha}^{-1} \right) \left( z+{z}^{-1}
 \right) -3
\end{array}
\right)\;.
\end{equation}
Non-trivial solutions of (\ref{eqc}) correspond to when 
\begin{equation}\label{det_sys}
\text{det}({\bf C\rm}({z,}\alpha)+ \omega^2{\bf I\rm})={0}\;,
\end{equation}
and this leads to a bi-quadratic equation for $\alpha$ in the form:
\begin{eqnarray}\label{lam_bi_quad}
&&\left(\alpha+\alpha^{-1}\right)^2-4b(z)\left(\alpha+\alpha^{-1}\right)+4c(z)=0.
\end{eqnarray}
Returning to (\ref{usol_outside}), using (\ref{lam_sol}) and (\ref{eqc}), the linearly independent solutions with appropriate behaviour as ${\tt y}\to+\infty$ correspond to $\alpha=\lambda_1$ and $\alpha=\lambda_2$, which implies the general solution in the upper-lattice half space takes the form
\begin{equation}\label{eq_u_full_sol}
{\mathbf{u}}^{\rm F}_{{\tt y}}(z)= g_1(z)\alpha^{\tt y}{\bf p}(\alpha)|_{\alpha=\lambda_1(z)}+g_2(z)\alpha^{\tt y}{\bf p}(\alpha)|_{\alpha=\lambda_2(z)}
\end{equation}
where 
 $g_j$, $j=1,2$, are yet to be determined and ${\bf p}(\alpha)$ is stated in (\ref{BM}) as ${\bf c}$ with $\lambda$ replacing $\alpha$.
Setting ${{\tt y}}=0$, we then extend our representation (\ref{eq_u_full_sol}) to the lattice row containing defect. In doing so, we find the vector ${\bf g}(z)=(g_1(z), g_2(z))^T$ takes the form 
\[{\bf d}={\bf B}^{-1}{\mathbf{u}}_0^{{\rm F}}\]
with the coefficient matrix in the right-hand side defined in (\ref{BM}). 

To complete the derivation of (\ref{u_upper}), we should derive the solution to (\ref{bulk_eq_tran}) in the lower lattice half-plane (i.e. for ${\tt y}\in{{\mathbb{Z}}_-}$).
We note that in addition to the roots $\lambda_j$, $j=1,2,$ of (\ref{lam_bi_quad}), there are also the solutions
\begin{equation}\label{mod_ge_1}
\lambda_3=\lambda_1^{-1} \quad \text{ and } \quad \lambda_4=\lambda_2^{-1}\quad \text{ with }\quad |\lambda_{3,4}|> 1\quad\text{ if } \text{Im}(\omega)>0\;.
\end{equation}
These roots are needed for the construction of the transformed scattered field for ${\tt y}\in{{\mathbb{Z}}_-}$ as described below.
Thus, for ${{\tt y}}\in{{\mathbb{Z}}_-}$, in contrast to \eqref{usol_outside}, the transformed scattered field can be determined using the representation
\begin{equation}\label{usol_outside_2}
{\mathbf{u}}^{\rm F}_{{\tt y}}={\bf q}(\alpha) \alpha^{\tt y}\;, \quad \text{ with } |\alpha| > 1\;, \quad {{{\tt y}} \in{{\mathbb{Z}}_-}}\;,
\end{equation}
and ${\bf q}(\alpha)$ is an eigenvector corresponding to $\alpha$ to be found.
One can check that this leads again to (\ref{det_sys}).
It then follows that (\ref{mod_ge_1}) and the constraint imposed on the magnitude of $\alpha$, that this parameter should take the values
$\alpha=\lambda_3=\lambda_1^{-1} \text{ and }\alpha=\lambda_4=\lambda_2^{-1}$ for the respective linearly independent solutions.
Additionally, note that eigenvectors ${\bf q}$ in (\ref{usol_outside_2}) corresponding to these choices are
\[{\bf q}(\alpha)|_{\alpha=\lambda_j^{-1}}={\bf J}{\bf p}(\lambda_j)\;, \quad j=1,2\;, 
\]
where ${{\textbf J}}$ is given by \eqref{BM} and ${\bf p}(\lambda_j)$ same as those from \eqref{eq_u_full_sol}.
Hence for ${{\tt y}}\in{{\mathbb{Z}}_-}$, we have 
\[{\mathbf{u}}^{\rm F}_{{\tt y}}(z)=f_1(z){\bf J}{\bf p}(\lambda_1(z))\lambda_1^{-{\tt y} }(z)+f_2(z){\bf J}{\bf p}(\lambda_2(z))\lambda_2^{-{\tt y} }(z)\;,\]
where the $f_j$, $j=1,2,$ can be found as a function of $z$ by taking ${{\tt y}}=0$, as before in case of ${\tt y}\in{{\mathbb{Z}}_+\setminus\{0\}}$. 
This completes the derivation of (\ref{u_upper}).

\section{Derivation of the Wiener-Hopf equation} 
\label{appWH}
In this section we derive the Wiener-Hopf equation (\ref{EqWH1}).
We begin by taking (\ref{eq2bSC}) and applying the transform (\ref{ZT}). This leads to an equation for the scattered field along ${\tt y}=0$ given by
\begin{eqnarray}
 -\omega^2{\mathbf{u}}^{\rm F}_{0}-\beta \omega^2{\mathbf{u}}_{0-}-\beta \omega^2[{{\bf a}}^{\mathfrak{i}}e^{\text{i}k_{\tt x}{\tt x}}]_-
&=& {\bf a}^{(1)}\cdot {\mathbf{u}}^{{\rm F}}_{0}\left[z^2+z^{-2}-2\right]{\bf a}^{(1)}\nonumber \\
&&+{\bf a}^{(2)}\cdot \left[{\mathbf{u}}^{{\rm F}}_{1}z^{-1}+{\mathbf{u}}^{{\rm F}}_{-1}z-2{\mathbf{u}}^{{\rm F}}_{0}\right]{\bf a}^{(2)}\nonumber\\
&&+{\bf a}^{(3)}\cdot \left[{\mathbf{u}}^{{\rm F}}_{1}z+{\mathbf{u}}^{{\rm F}}_{-1} z^{-1}-2{\mathbf{u}}^{{\rm F}}_{0}\right]{\bf a}^{(3)}\;.\nonumber\\
\label{defecteq1}
\end{eqnarray}
{
In going forward, it is convenient to use the following representation concerning the transformed displacement of the scattered wave:
\begin{equation}\label{amp_rep}
{{{\mathbf{u}}}}^{{\rm F}}_{{\tt y}}=\left\{
\begin{array}{ll}
{{{\bf W}}}^{{\rm F}}_{{\tt y}},&\quad \text{ if } {\tt y}\in{{\mathbb{Z}}_+}\;, \\[1ex]
{{{\bf V}}}^{{\rm F}}_{{\tt y}}, &\quad \text{ if } {\tt y}\in{{\mathbb{Z}}_-}\cup\{0\},
\end{array}\right.
\end{equation}
where ${{{\bf W}}}^{{\rm F}}_{{\tt y}}$ and ${{{\bf V}}}^{{\rm F}}_{{\tt y}}$ are given by the right-hand side of (\ref{u_upper}) for ${\tt y} \in{{\mathbb{Z}}_+}$ and ${\tt y}\in{{\mathbb{Z}}_-}$, respectively.}
Then, we have (\ref{defecteq1}) can be written as:
\begin{eqnarray}
 -\omega^2{\bf W}^{\rm F}_{0}-\beta \omega^2{\mathbf{u}}_{0-}-\beta \omega^2[{{\bf a}}^{\mathfrak{i}}e^{\text{i}k_{\tt x}{\tt x}}]_-
&=& {\bf a}^{(1)}\cdot {\bf W}^{{\rm F}}_{0}\left[z^2+z^{-2}-2\right]{\bf a}^{(1)}\nonumber \\
&&+{\bf a}^{(2)}\cdot \left[{\bf W}^{{\rm F}}_{1}z^{-1}+{\bf V}^{{\rm F}}_{-1}z-2{\bf W}^{{\rm F}}_{0}\right]{\bf a}^{(2)}\nonumber\\
&&+{\bf a}^{(3)}\cdot \left[{\bf W}^{{\rm F}}_{1}z+{\bf V}^{{\rm F}}_{-1} z^{-1}-2{\bf W}^{{\rm F}}_{0}\right]{\bf a}^{(3)}\;.\nonumber\\
\label{defecteq2}
\end{eqnarray}
Note that by construction, the representations in \eqref{amp_rep} satisfy (\ref{bulk_eq_tran}) for any $n_2\in{{\mathbb{Z}}\setminus\{0\}}$. 
Hence, we can exploit this to rewrite (\ref{defecteq2}) in the more compact form:
\begin{eqnarray}
&&-\beta \omega^2{\mathbf{u}}_{0-}-\beta \omega^2[{{\bf a}}^{\mathfrak{i}}e^{\text{i}k_{\tt x}{\tt x}}]_-\nonumber \\
&=& {\bf a}^{(2)}\cdot [{\bf V}^{{\rm F}}_{-1}-{\bf W}^{{\rm F}}_{-1}]z{\bf a}^{(2)}+{\bf a}^{(3)}\cdot [{\bf V}^{{\rm F}}_{-1}-{\bf W}^{{\rm F}}_{-1} ]z^{-1}{\bf a}^{(3)}\;.\end{eqnarray}
The right-hand sides of (\ref{u_upper}) then give
\begin{eqnarray}
-\beta \omega^2{\mathbf{u}}_{0-}-\beta \omega^2[{{\bf a}}^{\mathfrak{i}}e^{\text{i}k_{\tt x}{\tt x}}]_-= z ({\bf a}^{(2)}\cdot ({{{\textbf J}}}{\bf B \rm}\Lambda {\bf B}^{-1}{{\textbf J}} -{\bf B \rm}\Lambda^{-1} {\bf B}^{-1})\mathbf{u}^{\rm F}_0){\bf a}^{(2)}\notag\\
+z^{-1}({\bf a}^{(3)}\cdot ({{{\textbf J}}}{\bf B \rm}\Lambda {\bf B}^{-1}{{\textbf J}} -{\bf B \rm}\Lambda^{-1} {\bf B}^{-1})\mathbf{u}^{\rm F}_0){\bf a}^{(3)}.
\label{wiener1}
\end{eqnarray}
Next, by simply rearranging (\ref{wiener1}), applying (\ref{splitpm}) we obtain after some simplification (\ref{EqWH1}).

\section{Solution of the Wiener-Hopf equations}
\label{appWHsol}
We discuss the procedure to solve (\ref{EqWH1}) and demonstrate how to obtain (\ref{SOL}). Firstly, note the function $\mathrm{f}_-$ in (\ref{Fm}) is analytic for $|z|<|e^{{\rm i}k_{\tt x}}|$. Note the circle $|z|=|e^{{\rm i}k_{\tt x}}|<1$ {(see the sentence just before Section \ref{secexactsol})} is contained inside or outside the unit circle in the complex plane depending on the sign $\text{Re}(k_{{\tt x}})$.
In either case, we can write
\begin{equation}\label{Fminus}
\mathrm{f}_-(z)=\mathrm{f}_-^{(1)}(z)+\mathrm{f}_-^{(2)}(z)\qquad \text{ for } |z|<|e^{{\rm i}k_{\tt x}}|,
\end{equation}
where $\mathrm{f}_-^{(j)}(z)$ are defined in (\ref{eqFj}).
According to (\ref{eqFj}), the right-hand side in (\ref{Fminus}) has simple poles at $z=w_j=(-1)^{j} e^{{\rm i}k_{\tt x}}$, $j=1,2$.

In addition, the matrix kernel of (\ref{EqWH1}) (see also (\ref{ML})) can be factorised according to 
(\ref{Fact1}) and (\ref{Fact2}). Using this (\ref{Fminus}) allows 
(\ref{EqWH1}) to be written as follows 
\[[{\bf L}_+(z)]^{-1} {\mathbf{u}}_{0+}+{\bf L}_-(z){\mathbf{u}}_{0-}={\bf f}_+(z)+{\bf f}_-(z)\]
which utilises the additive split in the right-hand side comprised of 
\[{\bf f}_+(z)=\sum_{j=1}^2 ([{\bf L}_+(z)]^{-1}-[{\bf L}_+(w_j)]^{-1}){\bf a}^{\mathfrak{i}}\mathrm{f}^{(j)}_-(z)\;,\]
and 
\[{\bf f}_-(z)=\sum_{j=1}^2 ([{\bf L}_+(w_j)]^{-1}-{\bf L}_-(z)){\bf a}^{\mathfrak{i}}\mathrm{f}^{(j)}_-(z)\;.\]
Then 
\[{\mathbf{u}}_{0+}={\bf L}_+(z){\bf f}_+(z) \qquad \text{ and } {\mathbf{u}}_{0-}=[{\bf L}_-(z)]^{-1}{\bf f}_-(z)\]
and upon combining these through (\ref{splitpm}), we receive (\ref{SOL}).

{\section{Reflected and transmitted components of the incident wave}\label{appbulk}
The modes ${\bf u}_{\tt x,y}^{\rm bulk}$ in (\ref{mode_rep}) corresponding to the reflected and transmitted components of the incident field are computed in a similar way to the localised defect modes in (\ref{ubehind}), except for the fact as $z\to w_j$, 
\begin{equation}\label{asyu0wj}
{\bf u}^{\rm F}_0 \sim {\bf G}_j\Big(1-\frac{z}{w_j}\Big)^{-1}
\end{equation}
with 
\begin{equation}\label{Gj}
{\bf G}_j=2^{-1}([{\bf I}-{\bf L}(w_j)]^{-1}) {\bf a}_{\frak{in}}\;,
\end{equation}
for $j=1,2.$
Thus,
\begin{equation}\label{ubulk}
{\mathbf{u}}^{\text{bulk}}_{{\tt x}, {\tt y}}(w_j)={\bf T}_{{\tt y}}^{(j)}w_j^{{\tt x}}H(-{\tt x})
\end{equation}
where
\begin{equation}\label{Tj}
{\bf T}^{(j)}_{ {\tt y}}=\left\{\begin{array}{ll}
\displaystyle{ {\bf P}({\tt y}){\bf G}_j}, &\qquad \text{ for } {\tt y}\in{{\mathbb{Z}}_+},
\\[1ex]
\displaystyle{{\bf Q}(-{\tt y}){\bf G}_j}, &\qquad \text{ for } {\tt y}\in{{\mathbb{Z}}_-}\;.
\end{array}\right.
\end{equation}
An example of such a mode ${\mathbf{u}}^{\text{bulk}}_{{\tt x}, {\tt y}}$ is provided in Figure \ref{pic_mode_w1} of Section \ref{symmsec} for the case $z=w_2$. We report that the mode for $z=w_1$ appears to be very similar. There, unlike in Figure \ref{pic_mode_ahead_2}, the behaviour of the mode for ${\tt y}\ge 0$ and ${\tt y}<0$ is very distinct. This is due to the non-trivial coupling of the columns of ${\bf P}({\tt y})=\overline{{\bf Q}({\tt y})}$ (having real diagonal and imaginary off-diagonal components) brought by ${\bf G}_1=(-12+22{\rm i},0.1-15{\rm i})^{\rm T}\times 10^{-3}$ in (\ref{Gj}). Further, $|\lambda_j(w_2)|=1$, $j=1,2$, which physically means this mode is not localised to the defect in the direction perpendicular to it (compare with Figure \ref{pic_mode_ahead_2} where $|\lambda_j|<1$, $j=1,2$).}

\section{Dispersion relations for scattering response of the lattice}\label{singular}
Below we provide the dispersion relations corresponding to singular branch points and zeros of the functions $\mathfrak{L}_j(z)$, $j=1,2$.

\subsection{Dispersion relation for the remote mixed modes}\label{Singular1}
We begin with dispersion relation connected with appearance of the mixed mode vibrations discussed in Section \ref{symmsec}. This dispersion relation corresponds to the singular points belonging to both kernel functions $\mathfrak{L}_1(z)$ and $\mathfrak{L}_2(z)$ and provides the curves marked $``\mathcal{M}"$ in Figure \ref{pic_dd_bulk}. 
At these singular points $\lambda_1={\lambda_2}^{-1}$ with $|\lambda_j|=1$, $j=1,2$. This condition is satisfied at a subcollection of the twelve roots of:
\begin{eqnarray}
0&=&(z+1/z)^6+\frac{4(2\omega^2-9)}{3}(z+1/z)^4 \nonumber\\
&&+\frac{16((\omega^2-6)^2-9)}{9}(z+1/z)^2
-\frac{16((\omega^2-4)^2-4)}{3}\;.\label{eqrootconj}
\end{eqnarray}
As discussed in Section \ref{sec4}, the solutions of the preceding equation and the corresponding mixed modes appear for narrow ranges of the radian frequency within $\omega\in (0,\sqrt{6})$.

\subsection{Dispersion relation for dynamic shear and tensile modes in the bulk lattice}\label{Singular2}
Next, we present the dispersion relations defining the shear and tensile modes appearing in the remote lattice during the scattering of the incident wave. 

The tensile modes are connected with the singular points belonging exclusively to $\mathfrak{L}_1(z)$ and occur at:
\begin{equation}\label{z3pmA}
z=\pm z_-^{(1, \pm)}, \pm (z_-^{(1, \pm)})^{-1}, \qquad z_-^{(1, \pm)}=\frac{\sqrt{q_{\pm}+1}-\sqrt{q_{\pm}-1}}{\sqrt{q_{\pm}+1}+\sqrt{q_{\pm}-1}}\;, 
\end{equation}
with 
\[ q_{\pm}=\frac{1}{8}[-1\pm \sqrt{1-16(\omega^2-5)}]\;.\]
At these points, $\lambda_1=1$ ($\lambda_2=-1$) if $z= (z_-^{(1, \pm)})^{\pm 1}$ ($z=- (z_-^{(1, \pm)})^{\pm 1}$). 
The above points are connected with the dispersion curves marked ``$\mathcal{T}$" in Figure \ref{pic_dd_bulk}.

On the other hand, the shear modes correspond to singular branch points exclusive to the 
kernel function $\mathfrak{L}_2(z)$ and identify with the dispersion curves marked ``$\mathcal{S}$" in Figure \ref{pic_dd_bulk}. {Those branch points are obtained as solutions of the equations:
\begin{equation}\label{zOmpm}
z^2-2\Omega_{\pm} z+1=0
\end{equation}
with
\begin{equation}\label{OMpmeq}
\Omega_{\pm}=\pm \frac{\omega^2-3}{3}\;.
\end{equation}}
Thus, these points occur 
when
\begin{equation}
z=z_-^{(2)}, (z_-^{(2)})^{-1}\;, \quad z^{(2)}_-=\frac{\sqrt{\Omega_++1}-\sqrt{\Omega_+-1}}{\sqrt{\Omega_++1}+\sqrt{\Omega_+-1}}\;, \quad 
\label{z1pm}
\end{equation}
for those points $\lambda_1=-1$ or $\lambda_2=-1$, dependent on the frequency $\omega$. The remaining branch points appear for 
\begin{equation}
z=z_-^{(3)}, (z_-^{(3)})^{-1}\;, \quad z^{(3)}_-=\frac{\sqrt{\Omega_-+1}-\sqrt{\Omega_--1}}{\sqrt{\Omega_-+1}+\sqrt{\Omega_--1}}\;, \quad 
\label{z2pm}
\end{equation}
 where either $\lambda_1=1$ or $\lambda_2=1$ 
for $\omega\in (0, \sqrt{6})$. 
In this frequency range, the singularities in (\ref{z1pm}) and (\ref{z2pm}) correspond to real wave numbers.
The points in (\ref{z1pm}) and (\ref{z2pm}) are also removable singularities of the function $\mathfrak{L}_1(z)$.

\subsection{Dispersion relations for dynamic modes along the defect} \label{zeros}

Next we describe the simple zeros of the functions $\mathfrak{L}_1(z)$ and $\mathfrak{L}_2(z)$, that give information about longitudinal and transverse wave modes supported by the inertial defect. 

\subsubsection{Dispersion relation for longitudinal defect modes}
The longitudinal modes provided shear-type dynamic deformations in the lattice. They correspond to 
simple zeros of $\mathfrak{L}_1(z)$ that are traced from an subcollection of roots to the polynomial:
\begin{eqnarray}
0&=&(z^2+1/z^2)^7+c_1(\omega,\beta)(z^2+1/z^2)^6+c_2(\omega,\beta)(z^2+1/z^2)^5\nonumber
\\
&&
+c_3(\omega,\beta)(z^2+1/z^2)^4
+c_4(\omega,\beta)(z^2+1/z^2)^3+c_5(\omega,\beta)(z^2+1/z^2)^2\nonumber
\\
&&+c_6(\omega,\beta)(z^2+1/z^2)+c_7(\omega,\beta)\label{poly1}
\end{eqnarray}
with the coefficients $c_j$, $1\le j \le 7$, being given by 
\begin{eqnarray*}
&&c_1(\omega,\beta)=\frac{1}{6}(40\omega^2-111)\;,\qquad c_2(\omega,\beta)=-\frac{2(9\beta^2-83)}{9}\omega^4-105\omega^2+\frac{2289}{16}\;,\\
&&c_3(\omega,\beta)=-\frac{4(18 \beta^2-61)}{9}\omega^6+\frac{43(129\beta^2-1427)}{258}\omega^4+\frac{1345}{4}\omega^2-\frac{4805}{8}\;,\\
&&c_4(\omega,\beta)=\frac{(9\beta^2-55)^2-1216}{81}\omega^8+\frac{(591\beta^2-2432)}{9}\omega^6
\\&&\qquad\qquad-\frac{(3069\beta^2-42980)}{36}\omega^4-\frac{6740}{3}\omega^2+\frac{2965}{2}\;,\\
&&c_5(\omega,\beta)=\frac{[(6\beta^2-19)^2-97]}{27}\omega^{10}-\frac{2((18\beta^2-311)^2-71413)}{324}\omega^8\\
&& -\frac{8(207\beta^2-1096)}{9}\omega^6+\frac{(2799\beta^2-52594)}{18}\omega^4+4140\omega^2-2157\;,\\
&&c_6(\omega,\beta)=\frac{4(\beta^2-1)(\beta^2-4)}{9}\omega^{12}+\frac{(232\beta^2-352)}{9}\omega^{10}\\
&&-\frac{4((9\beta^2+125)^2-22546)}{81}\omega^8+\frac{4(459\beta^2-3388)}{9}\omega^6-(127\beta^2-3496)\omega^4\\
&&-4000\omega^2+1717\;,\\
\end{eqnarray*}
and
\begin{eqnarray*}
&&c_7(\omega,\beta)=\frac{8(\beta^2-1)(\beta^2+2)}{9}\omega^{12}-\frac{16((\beta^2+1)^2-7)}{3}\omega^{10}
+\frac{2((18\beta^2+49)^2-11725)}{81}\omega^8\\
&&-\frac{16(13\beta^2-158)}{3}\omega^6+\frac{2(153\beta^2-7382)}{9}\omega^4+\frac{4760}{3}\omega^2-578\;.
\end{eqnarray*}

\subsubsection{Dispersion relation for transverse defect modes}
The transverse vibration modes of the defect promote tensile-type modes in the lattice and 
corresponding to an admissible subset of the roots of:
\begin{eqnarray}
0&=&(z^2+1/z^2)^5+d_1(\omega,\beta)(z^2+1/z^2)^4+d_2(\omega,\beta)(z^2+1/z^2)^3\nonumber\\
&&+d_3(\omega,\beta)(z^2+1/z^2)^2+d_4(\omega,\beta)(z^2+1/z^2)+d_5(\omega,\beta) \label{poly2}
\end{eqnarray}
where the coefficients $d_j$, $1\le j \le 5$, are 
\begin{eqnarray*}
&&d_1(\omega,\beta)=\frac{8(\beta^2-1)}{9}\omega^4+8\omega^2-10\;,\\
&&d_2(\omega,\beta)=\frac{16(\beta^2-1)^2}{81}\omega^8+\frac{16(7\beta^2-8)}{27}\omega^6-\frac{4(13\beta^2-68)}{9}\omega^4-64\omega^2+40\;,\\
&&d_3(\omega,\beta)=\frac{64(\beta^2-1)(\beta^2-2)}{243}\omega^{10}-\frac{8[(2\beta^2-19)^2-269]}{81}\omega^8-\frac{512(\beta^2-3)}{27}\omega^6\\
&&\qquad \qquad +\frac{8(15\beta^2-178)}{9}\omega^4+192\omega^2-80\;,\\
&&d_4(\omega,\beta)=\frac{64(\beta^2-1)(\beta^2-4)}{729}\omega^{12}+\frac{128(3\beta^2-4)}{81}\omega^{10}-\frac{64[(\beta^2+11)^2-178]}{81}\omega^8\\
\\&&\qquad \qquad+\frac{64(11\beta^2-68)}{27}\omega^6-\frac{112(\beta^2-24)}{9}\omega^4-256\omega^2+80\;,
\end{eqnarray*}
and
\begin{eqnarray*}
&&d_5(\omega,\beta)=\frac{128(\beta^2-1)(\beta^2+2)}{729}\omega^{12}-\frac{768[(\beta^2+1)^2-7]}{729}\omega^{10}+\frac{32[(2\beta^2+5)^2-133]}{81}\omega^8\\
&&\qquad\qquad -\frac{256(\beta^2-14)}{27}\omega^6+\frac{32(\beta^2-54)}{9}\omega^4+128\omega^2-32\;.
\end{eqnarray*}

{
\section{Results used in the evaluation of contributions to the far-field lattice response}
\subsection{Formal justification of asymptotes of ${\bf u}_0^{\rm F}$ near singular branch points}\label{secF1}
Here, we present a formal argument that the solution ${\bf u}_0^{\rm F}$ has the asymptote (\ref{asympbranch}) for $z\to z_b$, $z\in \mathcal{B}$ with $|z_b|<1$.
We show this for the case where $z_b=z_-^{(2)}$, given in (\ref{z1pm}). 
We note that according to (\ref{zOmpm}) that 
\[\Omega_+=\frac{1}{2}\left(z_b+\frac{1}{z_b}\right)\;,\]
from which it follows with (\ref{OMpmeq}) that 
\[\omega^2=\frac{3}{2}\left(z_b+\frac{1}{z_b}\right)+3\;,\]
where here and going forward $z_b=z_-^{(2)}$.
Substitution of this into the expressions in (\ref{lam_sol}) taking $z\to z_b$, one finds:
\[b(z)=-\frac{(z_b^2+1)(z_b^4+2z_b^3+2 z_b+1)}{4 z_b^3}+O(z-z_b)\;,\]
\[b(z)^2-c(z)=\frac{(z_b^4+1)^2(z_b+1)^4}{16 z_b^6}+O(z-z_b)\;.\]
Therefore, we can assert
\begin{eqnarray}\label{mjasymp}
m_j(z)&=&m^j_0+O(z-z_b)\;, \qquad z\to z_b
\end{eqnarray}
where 
\begin{eqnarray}
m_0^j&=&-\frac{(z_b^2+1)(z_b^4+2z_b^3+2 z_b+1)}{4 z_b^3}\nonumber \\
&&+(-1)^{j+1}\frac{(z_b^4+1)(z_b+1)^2}{4 z_b^3}S\left[{\rm Im}\left(\frac{(z_b^4+1)^2(z_b+1)^4}{16 z_b^6}\right)\right]\;.
\end{eqnarray}
for $j=1,2$, with
\[S[x]=\left\{\begin{array}{ll}
1, & \qquad x\ge 0 \\
0, & \qquad \text{otherwise}\;.
\end{array}\right.\]
Here, we note
\[m_0^j=-1 \qquad \text{ if }\qquad (-1)^{j+1}S\left[{\rm Im}\left(\frac{(z_b^4+1)^2(z_b+1)^4}{16 z_b^6}\right)\right]=1 \;, j=1,2.
\]
Hence, to leading order, only $m_1(z)$ or $m_2(z)$ is equal to $-1$.
In such a case, insertion of the corresponding asymptote from (\ref{mjasymp}) in $\lambda_j(z)$ in (\ref{lam_sol}) gives
\begin{eqnarray*}
\lambda_{j}(z)&=&\frac{O((z-z_b)^{1/2})-\sqrt{-2+O((z-z_b)^{1/2})}}{{O((z-z_b)^{1/2})}+\sqrt{-2+O((z-z_b)^{1/2})}}\\
&=&-1+O((z-z_b)^{1/2})
\end{eqnarray*}
for $j=1$ or 2.
This together with (\ref{zOmpm}) then enables us to show $\mathfrak{L}_j(z)$, $j=1,2$, in (\ref{L1L2}) have the asymptotics:
\[\mathfrak{L}_1(z)= 1+O((z-z_b)^{1/2}),\quad
\mathfrak{L}_2(z)= \text{Const}\, (z-z_b)^{-1/2}\;,\]
for $z\to z_b$. Those asymptotes also persist for $z\to z_b$ in the matrix function ${\bf L}_+(z)$ which retains the singular properties of ${\bf L}(z)$ inside the unit disk and that is involved in the representation of ${\bf u}_0^{\rm F}$ in (\ref{SOL}). Hence, the asymptote specified in (\ref{asympbranch}) holds and above approach can be used to verify this asymptote for all $z\to z_b$ (with obvious modifications).

\subsection{Computation of (\ref{ointA})}\label{OintAsec}
Here, we show that
\begin{equation}\label{oint0}
\oint_{{\mathcal{C}}}z^{{\tt x}-1}\Big(1-\frac{z_b}{z}\Big)^{-1/2}\, {\rm d} z=2{\rm i}z_b^{\tt x}\,{\rm B}({\tt x}+1/2, 1/2)H({\tt x})\;,
\end{equation} 
which is used to evaluate the contribution of singular branch points at $z=z_b \in \mathcal{B}$ with $|z_b|<1$ in (\ref{UxyR}) when $|{\tt x}|\to \infty$ (see (\ref{mode_rep}). Here, the integrand on the above left-hand side has a linear branch cut connecting its zero branch point $z=0$ and singular branch point $z=z_b$ inside the unit circle $\mathcal{C}\subset \mathbb{C}$. We therefore evaluate the integral above using the contour $\tilde{\gamma}_1$ shown in Figure \ref{pic_contours_1}(a), i.e. we consider
\begin{equation}\label{eqoint1}\oint_{\tilde{{\gamma}_1}} z^{{\tt x}-1}\Big(1-\frac{z_b}{z}\Big)^{-1/2}\, {\rm d} z
\end{equation}
with $\mathcal{C}\subset \tilde{\gamma}_1$.
 The preceding integrand is analytic in $\tilde{\gamma}_1$, and therefore owing to Cauchy's theorem, we have 
\begin{equation}\label{refint1}
 0=\oint_{\tilde{{\gamma}}_1} z^{{\tt x}-1}\Big(1-\frac{z_b}{z}\Big)^{-1/2}\, {\rm d} z=z_b^{\tt x}\oint_{\gamma_r} w^{{\tt x}-1} \, \sqrt{\frac{w}{w-1}}\,{\rm d}{w}\;,
\end{equation}
where in moving to the second equality we employed the change of variable $w=z/z_b$, which transforms the contour of integration from $\tilde{\gamma}_1$ to $\gamma_r$ shown in Figure \ref{pic_contours_1}(b), with $r=|z_b|^{-1}>1$ governing the radius of the outer circular contour $\mathcal{C}_r\subset \gamma_r$.
\begin{figure}[!ht]
\center{\includegraphics[width=1\textwidth]{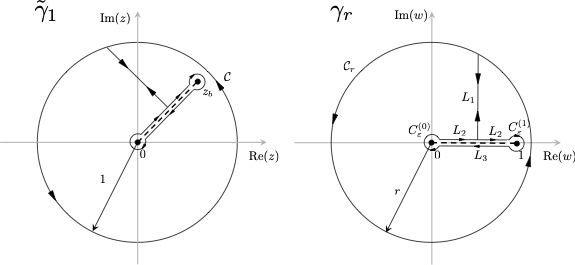}}
(a)~~~~~~~~~~~~~~~~~~~~~~~~~~~~~~~~~~~~~~~~~~~~~~~~~~~~~~~~~~~~(b)

\center{\includegraphics[width=0.5\textwidth]{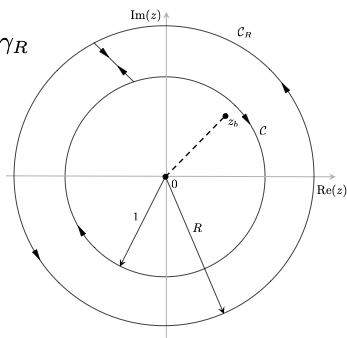}}

(c)
\caption{Contours of integration used in evaluating (\ref{oint0}). Orientation of the contours are indicated by arrows. The branch cuts are represented by dashed lines. }
\label{pic_contours_1}
\end{figure}
Next, let ${\tt x}>-1/2$. Note that
\[\oint_{\mathcal{C}^{(j)}_\varepsilon} w^{{\tt x}-1} \, \sqrt{\frac{w}{w-1}}\,{\rm d}{w}=O(\varepsilon^{{\tt x}+1/2})\]
for $j=1,2$, where $\varepsilon$ is chosen sufficiently small.
The preceding integrand is also analytic on $\mathfrak{L}_1$ of Figure \ref{pic_contours_1}(b). Hence, contour integrals over $\pm \mathfrak{L}_1$ cancel each other. Thus, in taking $\varepsilon \to +0$ we obtain:
\begin{eqnarray}
0=\oint_{\gamma_r}w^{{\tt x}-1} \, \sqrt{\frac{w}{w-1}}\,{\rm d}{w}&=& \left(\oint_{C_r}+\oint_{\mathfrak{L}_2}+\oint_{\mathfrak{L}_3}\right)w^{{\tt x}-1}\sqrt{\frac{w}{w-1}}\,{\rm d}{w}\nonumber\\
&=&\oint_{C_r} w^{{\tt x}-1} \, \sqrt{\frac{w}{w-1}}\,{\rm d}{w} + \left(\int_{0+{\rm i}0}^{1+{\rm i}0} +\int^{0-{\rm i}0}_{1-{\rm i}0}\right)w^{{\tt x}-1} \, \sqrt{\frac{w}{w-1}}\,{\rm d}{w}\nonumber\\ &=&\oint_{C_r} w^{{\tt x}-1} \, \sqrt{\frac{w}{w-1}}\,{\rm d}{w} -2{\rm i}\int_{0}^{1} w^{{\tt x}-1} \, \sqrt{\frac{w}{1-w}}\,{\rm d}{w}\label{eqoint2}
\end{eqnarray}
where the contour integrals over $\mathfrak{L}_2$ and $\mathfrak{L}_3$ where converted to integrals over the unit interval above and below the branch cut along $[0,1]$ and these integrals were then merged using the property of the principal branch of the square root.
Then, after recognising that
\[\int_{0}^{1} w^{{\tt x}-1} \, \sqrt{\frac{w}{1-w}}\,{\rm d}{w}={\rm B}({\tt x}+1/2, 1/2)\]
where $B(p,q)$ is the Beta function, and combining this with (\ref{eqoint1}), (\ref{refint1}) and (\ref{eqoint2}),
we obtain after returning all integrals to the form involving $z$, that 
\begin{equation}\label{oint1}
\oint_{{\mathcal{C}}} z^{{\tt x}-1}\Big(1-\frac{z_b}{z}\Big)^{-1/2}\, {\rm d} z=2{\rm i}z_b^{\tt x}\,{\rm B}({\tt x}+1/2, 1/2)\;,\qquad \text{ for }{\tt x}>-1/2 \;.
\end{equation}
If ${\tt x}\le -1/2$, one can compute the integral in (\ref{oint0}) with the contour $\gamma_R$ in Figure \ref{pic_contours_1}(c). Here, the integrand appearing in (\ref{oint0}) is analytic for $|z|>1$ and 
\[\int_{\mathcal{C}_R} z^{{\tt x}-1}\Big(1-\frac{z_b}{z}\Big)^{-1/2}\, {\rm d} z=O(R^{{\tt x }}).\]
Then, taking $R\to \infty$, allows one to show through Cauchy's theorem that 
\begin{equation}\label{oint2}
\oint_{{\mathcal{C}}} z^{{\tt x}-1}\Big(1-\frac{z_b}{z}\Big)^{-1/2}\, {\rm d} z=0\;,\qquad \text{ for }{\tt x}>-1/2\;.
\end{equation}
Thus, (\ref{oint1}) and (\ref{oint2}) gives the required result (\ref{oint0}).

\subsection{Evaluating the integral (\ref{ointsimppolesA})}
\label{secF3}
Next we show that 
\begin{equation}\label{ointsimppoles}
\oint_{\mathcal{C}}\frac{z^{{\tt x}-1}}{1-z z_0^{-1}}{\rm d} z=2\pi {\rm i}z_0^{\tt x}H(-{\tt x})\;,
\end{equation}
that contributes to the main asymptotics for the far-field lattice behaviour behind the defect tip. Here, $z_0\in \mathcal{Z}$, with $|z_0|>1$. We begin by considering the case ${\tt x}<1$ and the integral
\begin{equation}\label{z0int1}
\oint_{\sigma_R} \frac{z^{{\tt x}-1}}{1-z z_0^{-1}} {\rm d} z
\end{equation}
where $\sigma_R\subset \mathbb{C}$ is shown in Figure \ref{pic_contours_2}(a). There, the unit-disk $\mathcal{C} \subset \sigma_R$. Note that for the outer contour
\[\int_{\mathcal{C}_R} \frac{z^{{\tt x}-1}}{1-z z_0^{-1}}{\rm d} z=O(R^{{\tt x}-1})\]
Further, the integrand of (\ref{z0int1}) is analytic inside $\sigma_R$ except at $z=z_0$ where it has a simple pole. Thus, the line integrals over $\pm L$ will cancel each other. Further, taking $R\to \infty$ and applying the residue theorem establishes that
\[\lim_{R\to \infty}\oint_{\sigma_R} \frac{z^{{\tt x}-1}}{1-z z_0^{-1}}{\rm d} z=\oint_{-\mathcal{C}}\frac{z^{{\tt x}-1}}{1-z z_0^{-1}}{\rm d} z=-2\pi{\rm i} z_0^{\tt x}\]
Thus,
\begin{equation}\label{F2res1}
\oint_{\mathcal{C}}\frac{z^{{\tt x}-1}}{1-z z_0^{-1}}{\rm d} z=2\pi{\rm i} z_0^{\tt x}\;\qquad \text{ for }{\tt x}<1\;.
\end{equation}
The case when ${\tt x}\ge 1$ follows trivially from the application of Cauchy's theorem to the integral on the left-hand side of (\ref{ointsimppoles}) as the corresponding integrand is analytic inside the unit disk $\mathcal{C}$. 
\begin{figure}[!ht]
\center{\includegraphics[width=0.5\textwidth]{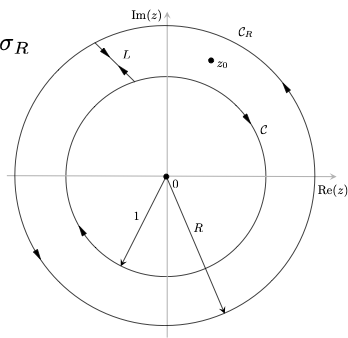}}
\caption{Contour of integration $\sigma_R$ used in the evaluation (\ref{ointsimppoles}) for ${\tt x}<1$. }
\label{pic_contours_2}
\end{figure}
Therefore, 
\begin{equation}
\oint_{\mathcal{C}}\frac{z^{{\tt x}-1}}{1-z z_0^{-1}}{\rm d} z=0\;\qquad \text{ for }{\tt x}\ge 1\;.
\label{F2res2}
\end{equation}
Then, (\ref{ointsimppoles}) follows from (\ref{F2res1}) and (\ref{F2res2}).
\subsection{Derivation of (\ref{mode_rep})}\label{App_mode_rep}
We take (\ref{UxyR}) and using (\ref{asympbranch}), (\ref{UxyRsimplpole}) and (\ref{asyu0wj}) rewrite it as:  
\[{\mathbf{u}}^{\mathfrak{s}}_ {{\tt x}, {\tt y}}=\sum_{j=1,2} {\mathbf{u}}^{\text{bulk}}_ {{\tt x}, {\tt y}}(w_j)+\sum_{z_b\in {\mathcal{B}}}{\mathbf{u}}^{\text{ahead}}_ {{\tt x}, {\tt y}}(z_b)+\sum_{z_0\in {\mathcal{Z}}}{\mathbf{u}}^{\text{behind}}_{{\tt x}, {\tt y}}(z_0)+{\mathbf{u}}^{\text{rem}}_ {{\tt x}, {\tt y}}\]
where 
\begin{eqnarray}
{\mathbf{u}}^{\text{bulk}}_ {{\tt x}, {\tt y}}(w_j)&=&\oint_{\mathcal{C}}{\bf T}_{\tt y}^{(j)}(1-zw_j^{-1})^{-1} \, {\rm d} z \label{intubulk}\\
{\mathbf{u}}^{\text{ahead}}_ {{\tt x}, {\tt y}}(w_j)&=&\oint_{\mathcal{C}}{\bf N}_{\tt y}(z_b)(1-z_b z^{-1})^{-1/2} \, {\rm d} z \label{intuahead}\\
{\mathbf{u}}^{\text{behind}}_ {{\tt x}, {\tt y}}(z_0)&=&\oint_{\mathcal{C}} {\bf S}_{\tt y}(z_0)(1-zz_0^{-1})^{-1} \, {\rm d} z \label{inubehind}
\end{eqnarray}
and
\begin{eqnarray*}
{\mathbf{u}}^{\text{rem}}_ {{\tt x}, {\tt y}}&=&\oint_{\mathcal{C}}\Big({\bf u}_{\tt y}^{\rm F} -\sum_{j=1}^2 {\bf T}_{\tt y}^{(j)}(1-zw_j^{-1})^{-1} -\sum_{z_0\in \mathcal{Z}}{\bf S}_{\tt y}(z_0)(1-zz_0^{-1})^{-1}\nonumber \\
&&  -\sum_{z_b\in\mathcal{B}}{\bf N}_{\tt y}(z_b)(1-z_b z^{-1})^{-1/2} \Big) z^{{\tt x}-1}{\rm d} z\;.
\end{eqnarray*}
(see (\ref{u_upper}), (\ref{SOL}), (\ref{Ny}), (\ref{Sy}) and (\ref{Tj}) for the definitions of terms appearing in the previous integrand). The integrals in (\ref{intubulk}), (\ref{intuahead}) and (\ref{inubehind}) are then evaluated with the results of Appendix \ref{OintAsec} and \ref{secF3}. This concludes the derivation of (\ref{mode_rep}).}

\end{document}